\documentclass[pdftex,twocolumn,3]{jour3}          

\usepackage[utf8]{inputenc}
\usepackage{epsf}
\usepackage{latexsym,amssymb,euscript}
\usepackage{widetext}
\usepackage[dvips]{graphicx}
\usepackage[numbers,sort&compress]{natbib}
\usepackage{amsmath}
\usepackage{nicefrac}
\usepackage{slashed}
\usepackage{booktabs}
\usepackage{hyperref}
\usepackage{braket}
\usepackage{chngcntr}
\usepackage{bm}
\usepackage{bbold}
\usepackage{graphics}
\usepackage{graphicx}
\usepackage{mciteplus}
\usepackage{bbold}
\usepackage{pdfpages}
\usepackage[titletoc]{appendix}
\graphicspath{{./figures/}}
\hypersetup{
 linktocpage = true,
 urlcolor = citegreen,
 colorlinks = true,
 linkcolor = citegreen,
 anchorcolor = citegreen,
 citecolor = purple,
 pdfstartview = {XYZ null null 1.25} 
           }
\usepackage[left=2cm, right=2cm]{geometry}
\usepackage{pstricks}
\usepackage{color}
\usepackage{xcolor}
\definecolor{urlblue}{rgb}{0.2,0.4,0.7}
\definecolor{citegreen}{rgb}{0,0.4,0.2}
\definecolor{linkred}{rgb}{0.9,0.2,0.1}
\usepackage{float}
\usepackage{academicons}
\definecolor{orcidlogocol}{HTML}{A6CE39}
\usepackage{fancyhdr}
\newcommand{\drv}{{\rm d}}

\newcommand{\tcite}[1]{~\cite{#1}}
\newcommand{\tref}[1]{~\ref{#1}}
\newcommand{\eref}[1]{~\eqref{#1}}

\newcommand{\tarr}{
\begin{array}}
\newcommand{\earr}{\end{array}}

\newcommand{\orcidADB}{\href{https://orcid.org/0000-0002-6114-7044}{\includegraphics[scale=0.1]{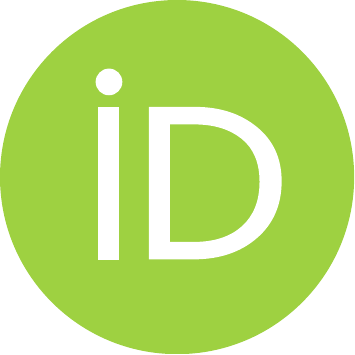}}}

\newcommand{\orcidFGC}{\href{https://orcid.org/0000-0003-3299-2203}{\includegraphics[scale=0.1]{figures/logo-orcid.pdf}}}

\newcommand{\orcidDYI}{\href{https://orcid.org/0000-0001-5701-4364}{\includegraphics[scale=0.1]{figures/logo-orcid.pdf}}}

\newcommand{\orcidAP}{\href{https://orcid.org/0000-0001-8984-3036}{\includegraphics[scale=0.1]{figures/logo-orcid.pdf}}}

\newcommand{\orcidWS}{\href{https://orcid.org/0000-0002-9764-3138}{\includegraphics[scale=0.1]{figures/logo-orcid.pdf}}}

\newcommand{\orcidAS}{\href{https://orcid.org/0000-0001-5247-8442}{\includegraphics[scale=0.1]{figures/logo-orcid.pdf}}}

\smartqed  

\journalname{}

\begin{document}

\title{Exclusive emissions of polarized $\rho$ mesons at the EIC and the proton content at low $x$
}

\subtitle{}

\author{
Andr\`ee Dafne Bolognino
\thanksref{e1,addr1,addr2} \orcidADB
\and
Francesco Giovanni Celiberto
\thanksref{e2,addr3,addr4,addr5} \orcidFGC
\and
Dmitry Yu. Ivanov
\thanksref{e3,addr6} \orcidDYI
\and
Alessandro Papa
\thanksref{e4,addr1,addr2} \orcidAP
\and \\
Wolfgang Sch\"afer
\thanksref{e5,addr7} \orcidWS
\and
Antoni Szczurek
\thanksref{e6,addr7,addr8} \orcidAS
}

\thankstext{e1}{{\it e-mail}:
\href{mailto:ad.bolognino@unical.it}{ad.bolognino@unical.it}}
\thankstext{e2}{{\it e-mail}:
\href{mailto:fceliberto@ectstar.eu}{fceliberto@ectstar.eu} (corresponding author)}
\thankstext{e3}{{\it e-mail}:
\href{mailto:d-ivanov@math.nsc.ru}{d-ivanov@math.nsc.ru}}
\thankstext{e4}{{\it e-mail}:
\href{mailto:alessandro.papa@fis.unical.it}{alessandro.papa@fis.unical.it}}
\thankstext{e5}{{\it e-mail}:
\href{mailto:wolfgang.schafer@ifj.edu.pl}{wolfgang.schafer@ifj.edu.pl}}
\thankstext{e6}{{\it e-mail}:
\href{mailto:antoni.szczurek@ifj.edu.pl}{antoni.szczurek@ifj.edu.pl}}

\institute{Dipartimento di Fisica, Universit\`a della Calabria, I-87036 Arcavacata di Rende, Cosenza, Italy\label{addr1}
\and
Istituto Nazionale di Fisica Nucleare, Gruppo collegato di Cosenza, I-87036 Arcavacata di Rende, Cosenza, Italy\label{addr2}
\and
European Centre for Theoretical Studies in Nuclear Physics and Related Areas (ECT*), I-38123 Villazzano, Trento, Italy\label{addr3}
\and
Fondazione Bruno Kessler (FBK),
I-38123 Povo, Trento, Italy\label{addr4}
\and
INFN-TIFPA Trento Institute of Fundamental Physics and Applications,
I-38123 Povo, Trento, Italy\label{addr5}
\and
Sobolev Institute of Mathematics, 630090 Novosibirsk, Russia\label{addr6}
\and
Institute of Nuclear Physics Polish Academy of Sciences, ul. Radzikowskiego 152, PL-31-342, Krak\'ow, Poland\label{addr7}
\and
College of Natural Sciences, Institute of Physics, University of Rzesz\'ow, ul. Pigonia 1, PL-35-310 Rzesz\'ow, Poland\label{addr8}
\vspace{0.75cm}
}

\date{\today}

\maketitle

\section*{Abstract}

We present a new study on helicity amplitudes and cross sections for the exclusive production of $\rho$ mesons at the EIC in high-energy factorization. In this framework the analytic expression of amplitudes takes the form of a convolution between an off-shell impact factor, depicting the ($\gamma^* \to \rho$) transition, and a nonperturbative density, known as Unintegrated Gluon Distribution (UGD) that encodes information about the proton structure at low $x$ and evolves according to the BFKL equation. We come out with an evidence that observables sensitive to the polarizations of the incoming virtual photon and of the emitted meson allow us to discriminate among different UGD models and to gather quantitative information on the proton content at high energies.
\vspace{0.50cm} \hrule
\vspace{0.50cm}
{
 \setlength{\parindent}{0pt}
 \textsc{Keywords}: \vspace{0.15cm} \\ QCD phenomenology \\ High-energy factorization \\ $\rho$ mesons \\ Low $x$ \\ UGD
}
\vspace{0.50cm} \hrule


\section{Hors d'{\oe}uvre}
\label{sec:introduction}

Shedding light on the proton structure via a multi-di\-men\-sion\-al imaging of their constituents is a frontier research field at new-generation colliding machines.
Fundamental questions about the inner dynamics of strong interactions are still far to be answered.
We mainly refer to proton mass and spin puzzles, whose resolution key relies upon a viewpoint stretch from the collinear-factorization description to a 3D tomographic vision, elegantly afforded by the \emph{transverse-momentum-dependent} (TMD) formalism (see Refs.\tcite{Collins:1981uk,Collins:2011zzd} and references therein).
However, a pure TMD-driven approach may be not e\-nou\-gh at low $x$. Here, large $\ln (1/x)$-type logarithms enter the perturbative expansion with a power increasing with the order of the strong coupling, and they need to be \emph{resummed} by all-order techniques.
The Balitsky-Fadin-Kuraev-Lipatov (BFKL) resummation~\cite{Fadin:1975cb,Kuraev:1976ge,Kuraev:1977fs,Balitsky:1978ic} allows us to exactly account for these large logarithms both in the leading (LL$x$) in the next-to-leading (NLL$x$) logarithmic approximation, namely including all contributions proportional to $\alpha_s^{n} \ln (1/x)^n$ and to $\alpha_s^{n+1} \ln (1/x)^n$, respectively.

A first kind of reactions that serve as probes for BFKL consists in the inclusive \emph{semi-hard} detection\tcite{Gribov:1983ivg} of two objects having large transverse momenta and being well separated in rapidity (see Ref.\tcite{Celiberto:2017ius} for recent applications). Here, a \emph{hybrid} high-energy and collinear factorization is established\tcite{Colferai:2010wu} (see also Refs.\tcite{Deak:2009xt,Deak:2018obv,Blanco:2020akb,vanHameren:2020rqt,vanHameren:2022mtk}) as a convolution of high-energy resummed partonic cross sections and collinear densities (or/and fragmentation functions, when final-state hadrons are identified).
Examples of these two-particle configurations include emissions of: light\tcite{Ducloue:2013bva,Caporale:2014gpa,Celiberto:2015yba,Celiberto:2015mpa,Celiberto:2016ygs,Celiberto:2016vhn,Caporale:2018qnm,Celiberto:2022gji} and heavy jets\tcite{Celiberto:2017nyx,Bolognino:2019ccd,Bolognino:2019yls,Bolognino:2021mrc,Maciula:2022bfv}, light\tcite{Bolognino:2018oth,Bolognino:2019yqj,Celiberto:2020rxb} and heavy hadrons\tcite{Boussarie:2017oae,Celiberto:2021dzy,Guiot:2021vnp,Celiberto:2021fdp,Celiberto:2022dyf,Celiberto:2022keu,Celiberto:2022kza}, Higgs bosons\tcite{Celiberto:2020tmb,Celiberto:2022zdg,Hentschinski:2020tbi,Celiberto:2022fgx} and forward Drell-Yan pairs\tcite{Motyka:2014lya,Brzeminski:2016lwh,Motyka:2016lta,Celiberto:2018muu} in association with a possible tag of backward jets\tcite{Golec-Biernat:2018kem}.

A second class of BFKL probes is represented by single-forward or single-central detections. Here we access the proton content via the \emph{unintegrated gluon distribution} (UGD), whose evolution at low $x$ is regulated by the BFKL Green's function. As a nonperturbative density, the UGD in not well known and distinct phenomenological models for it have been proposed so far. 
Starting from the information encoded in the UGD, low-$x$ improved collinear\tcite{Ball:2017otu,Bonvini:2019wxf} and TMD\tcite{Bacchetta:2020vty,Celiberto:2021zww,Bacchetta:2022nyv} distributions were determined.
The UGD has been extensively investigated through the inclusive deep inelastic scattering\tcite{Hentschinski:2012kr,Hentschinski:2013id} and the exclusive electro- or photo-production of forward vector mesons at HERA\tcite{Anikin:2009bf,Anikin:2011sa,Besse:2013muy,Bolognino:2018rhb,Bolognino:2018mlw,Bolognino:2019bko,Bolognino:2019pba,Celiberto:2019slj,Bautista:2016xnp,Garcia:2019tne,Hentschinski:2020yfm} and at the Electron-Ion Collider (EIC)\tcite{Bolognino:2021niq,Bolognino:2021gjm,Bolognino:2022uty,Celiberto:2022fam}.
Studies of central emissions of quarkonium states~\cite{Kniehl:2016sap,Cisek:2017ikn,Maciula:2018bex,Babiarz:2019mag,Babiarz:2020jkh,Babiarz:2020jhy,Babiarz:2020vep,Schafer:2021cat,Babiarz:2021uvm,Cisek:2022uqx} have the twofold advantages of unraveling the heavy-hadron production mechanisms and of probing kinematic corners at the frontier between TMD and high-energy factorization.

In this work we focus on the exclusive leptoproduction of $\rho$ mesons in high-energy factorization. By investigating polarized cross sections at EIC nominal energies\tcite{AbdulKhalek:2021gbh,Khalek:2022bzd,Hentschinski:2022xnd}, we highlight how these observables are able to discriminate among distinct models and parametrizations for the low-$x$ UGD.

\section{Theoretical setup}
\label{sec:theory}

We study the exclusive $\rho$-meson production at the EIC through the following subprocess
\begin{eqnarray}
\label{eq:subprocess}
 \gamma^*_{\lambda_i} (Q^2) \, p \; \to \; \rho_{\lambda_f} p \;,
\end{eqnarray}
where a photon having virtuality $Q^2$ and polarization $\lambda_i$ is absorbed by the proton and a $\rho$-meson with spin $\lambda_f$ is emitted. The two polarizations $\lambda_{i,f}$ can have values $0$ (longitudinal) or $\pm 1$ (transverse).
The stringent semi-hard scale hierarchy holds, $W^2 \gg Q^2 \gg \Lambda^2_{\rm QCD}$, with $W$ being the subprocess center-of-mass energy and $\Lambda_{\rm QCD}$ the QCD hadronization scale, leads to low-$x$ values, $x = Q^2/W^2$. The BFKL approach affords us a high-energy factorized expression for he\-li\-ci\-ty-dependent amplitudes
\begin{eqnarray}
\label{eq:amplitude}
 {\cal T}_{\lambda_i \lambda_f}(W^2, Q^2) &=& \frac{i W^2}{(2 \pi)^2} \int \frac{\drv^2 \boldsymbol{p}_T}{(\boldsymbol{p}_T^2)^2} 
 \\ \nonumber
 &\times&
 \Phi^{\gamma^*_{\lambda_i} \to \rho_{\lambda_f}}(\boldsymbol{p}_T^2, Q^2) \, {\cal G} (x, \boldsymbol{p}_T^2) \,.
\end{eqnarray}
In Eq.\eref{eq:amplitude} $\Phi^{\gamma^*_{\lambda_i} \to \rho_{\lambda_f}}(q^2, Q^2)$ stands for the off-shell impact factor that portrays the $\gamma^* \to \rho$ transition and embodies collinear distribution amplitudes (DAs, for more details see Section~2 of Ref.\tcite{Bolognino:2021niq}), whereas ${\cal G} (x, \boldsymbol{p}_T^2)$ is the BFKL UGD. 
Starting from helicity amplitudes, we build longitudinally ($L$) and transversally ($T$) polarized cross sections as
\begin{eqnarray}
\label{eq:sigma}
\sigma_{L,T}\,(\gamma^*\,p \rightarrow \rho\,p) = \frac{1}{16 \pi \, b(Q^2)}\left|\frac{{\cal T}_{00,11}(W^2, Q^2)}{W^2}\right|^2\,,
\end{eqnarray}
where $b(Q^2)$ is the \emph{diffraction slope}, for which we employ the following parametrization\tcite{Nemchik:1997xb}
\begin{equation}
\label{eq:slope_B_rho}
b(Q^2) = \beta_0 - \beta_1\,\log\left(\frac{Q^2+m_\rho^2}{m^2_{J/\psi}}\right)+\frac{\beta_2}{Q^2+m_\rho^2}\,,
\end{equation}
with $\beta_0 = 6.5$ GeV$^{-2}$, $\beta_1 = 1.2$  GeV$^{-2}$, and $\beta_2 = 1.6$.
In our phenomenological analysis we adopt the seven models for the UGD briefly introduced in Section~3 of Ref.\tcite{Bolognino:2021niq}.

\begin{figure*}[!t]

\centering

\includegraphics[width=0.38\textwidth]{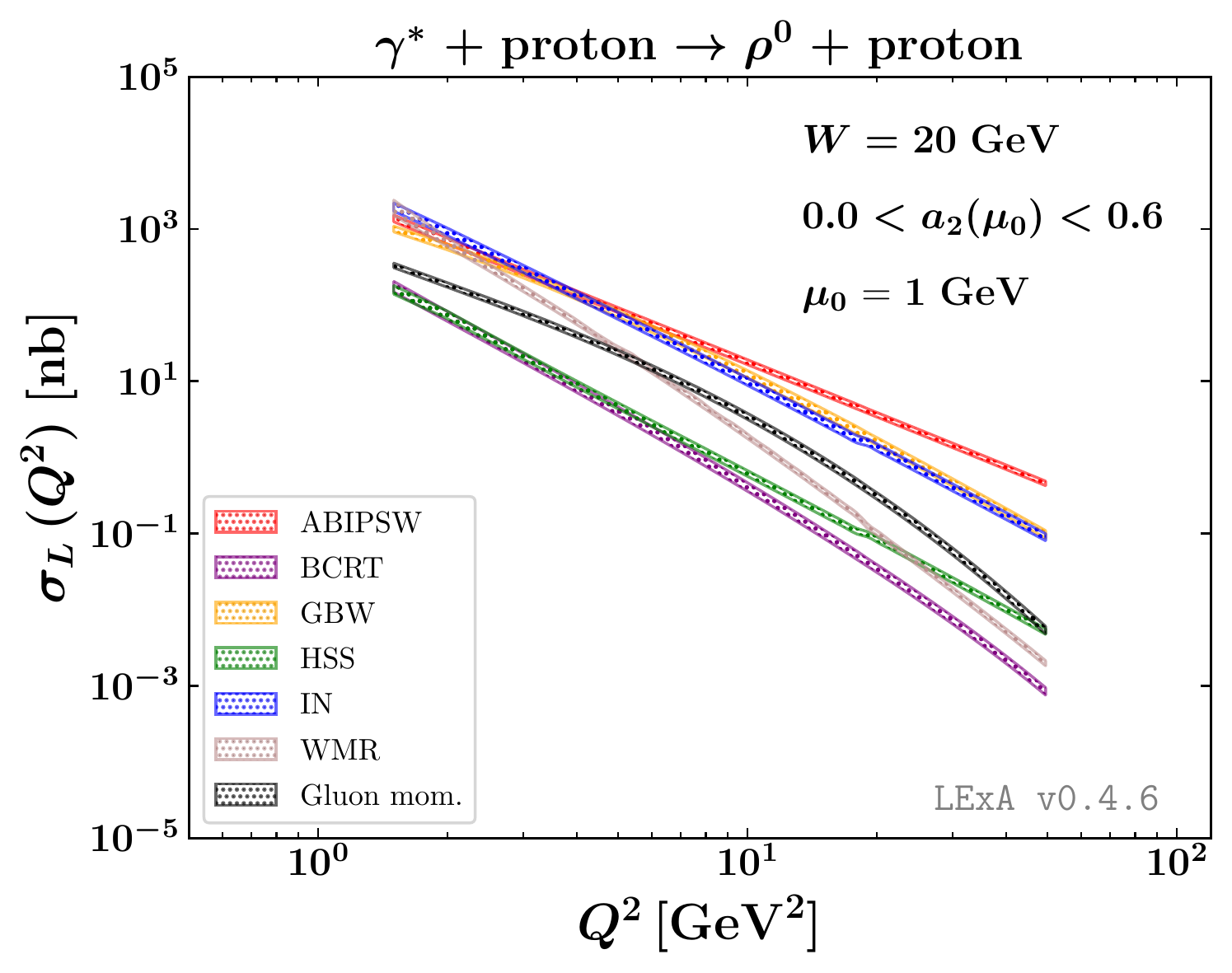} \hspace{0.50cm}
\includegraphics[width=0.38\textwidth]{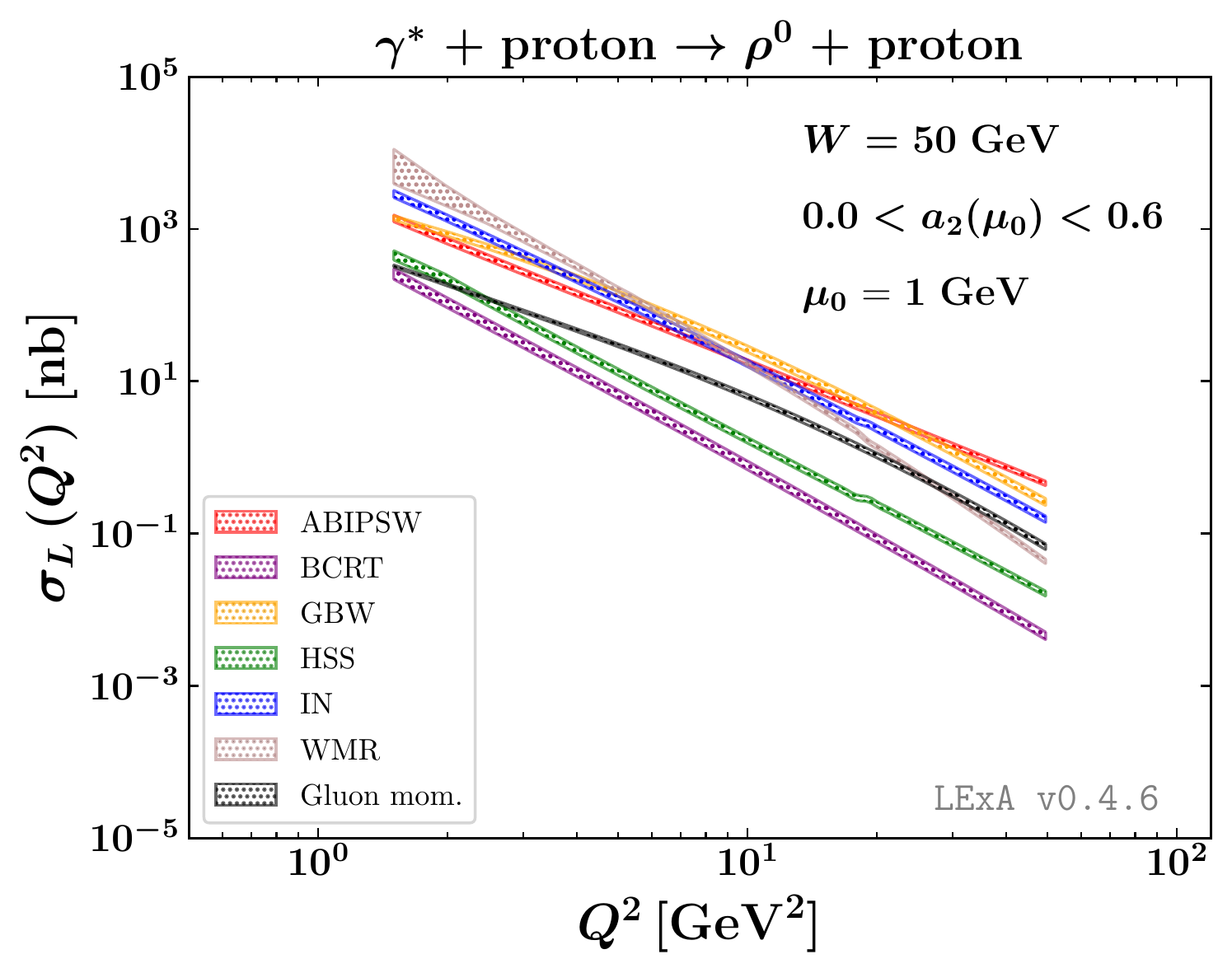}

\includegraphics[width=0.38\textwidth]{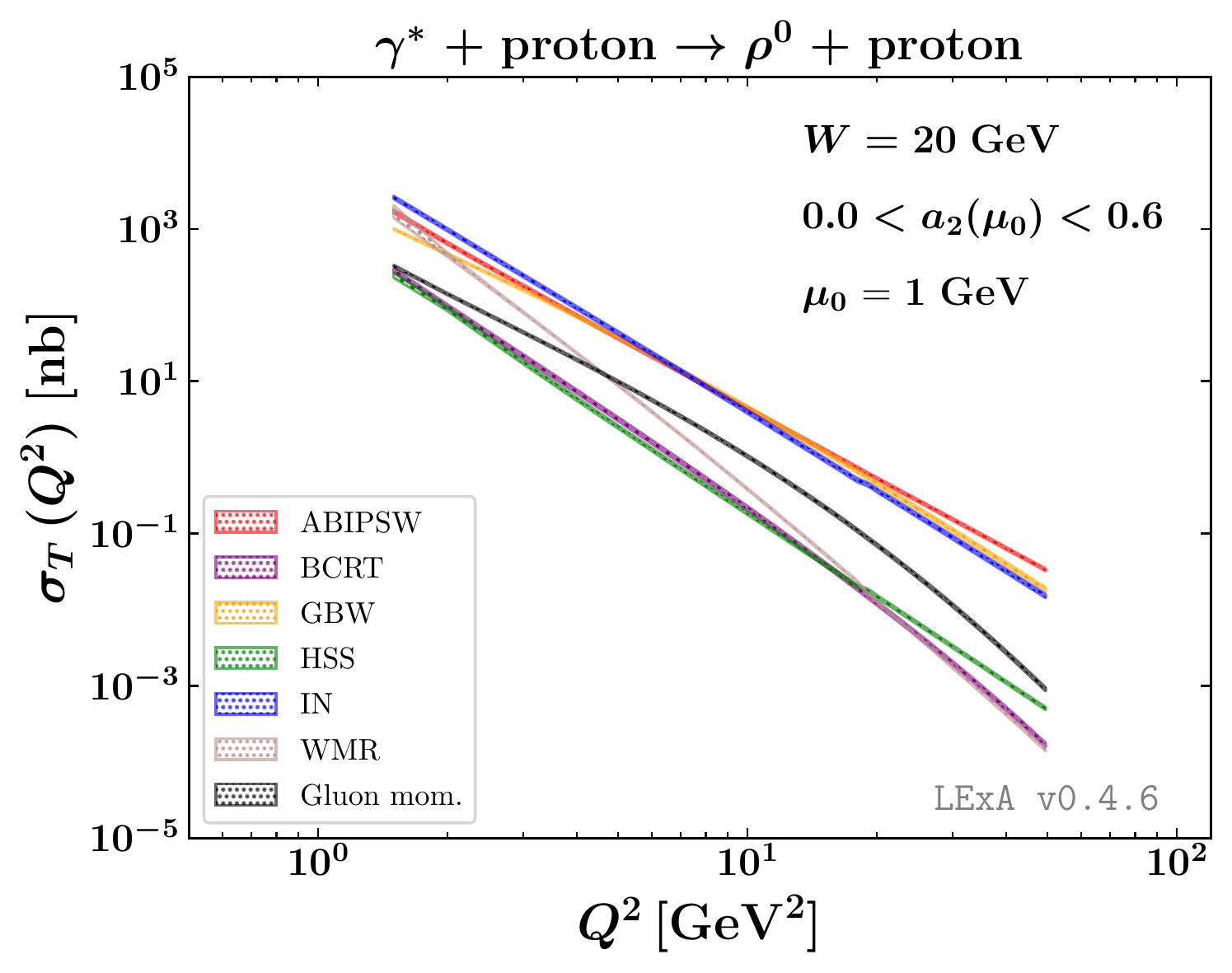} \hspace{0.50cm}
\includegraphics[width=0.38\textwidth]{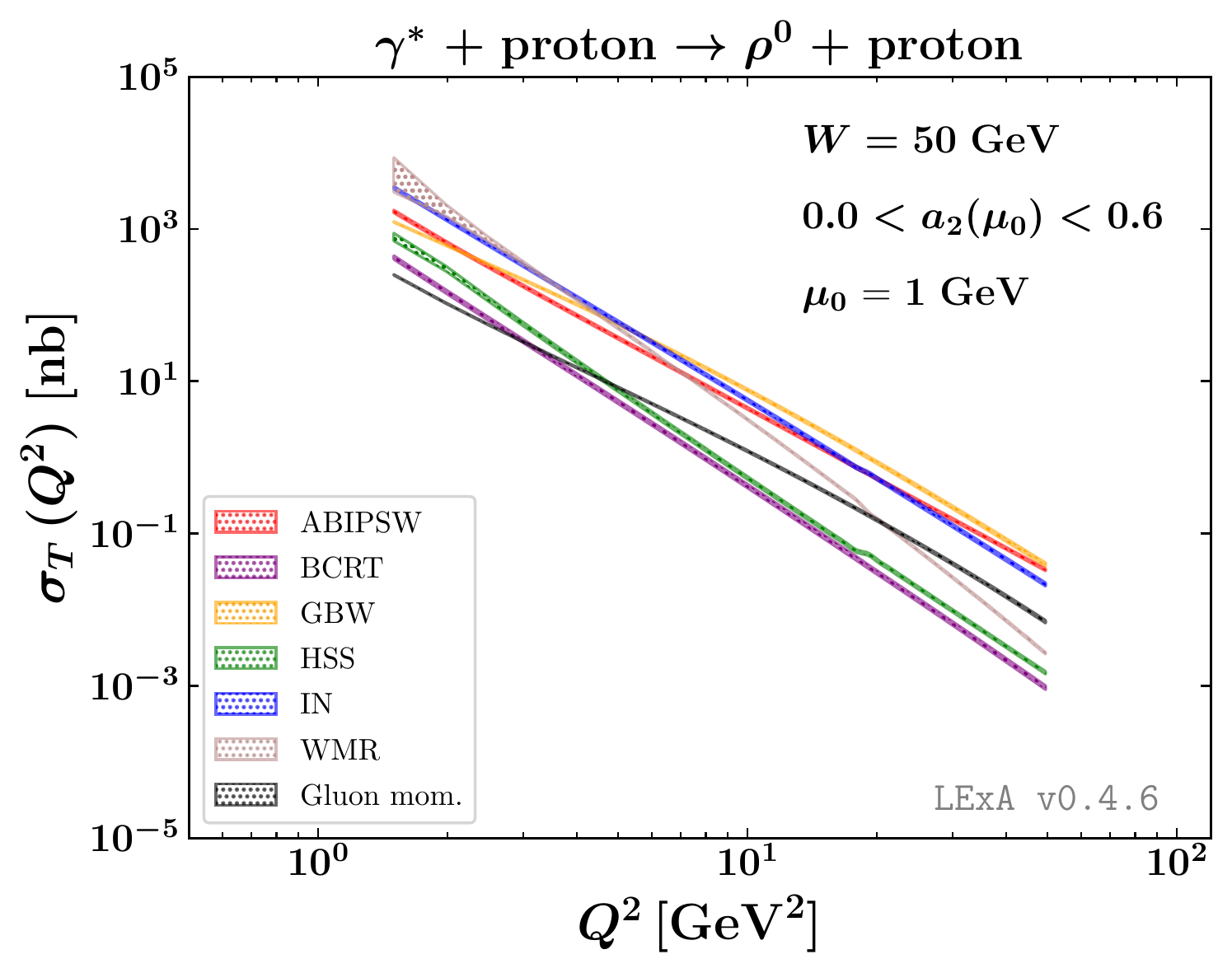}

\includegraphics[width=0.38\textwidth]{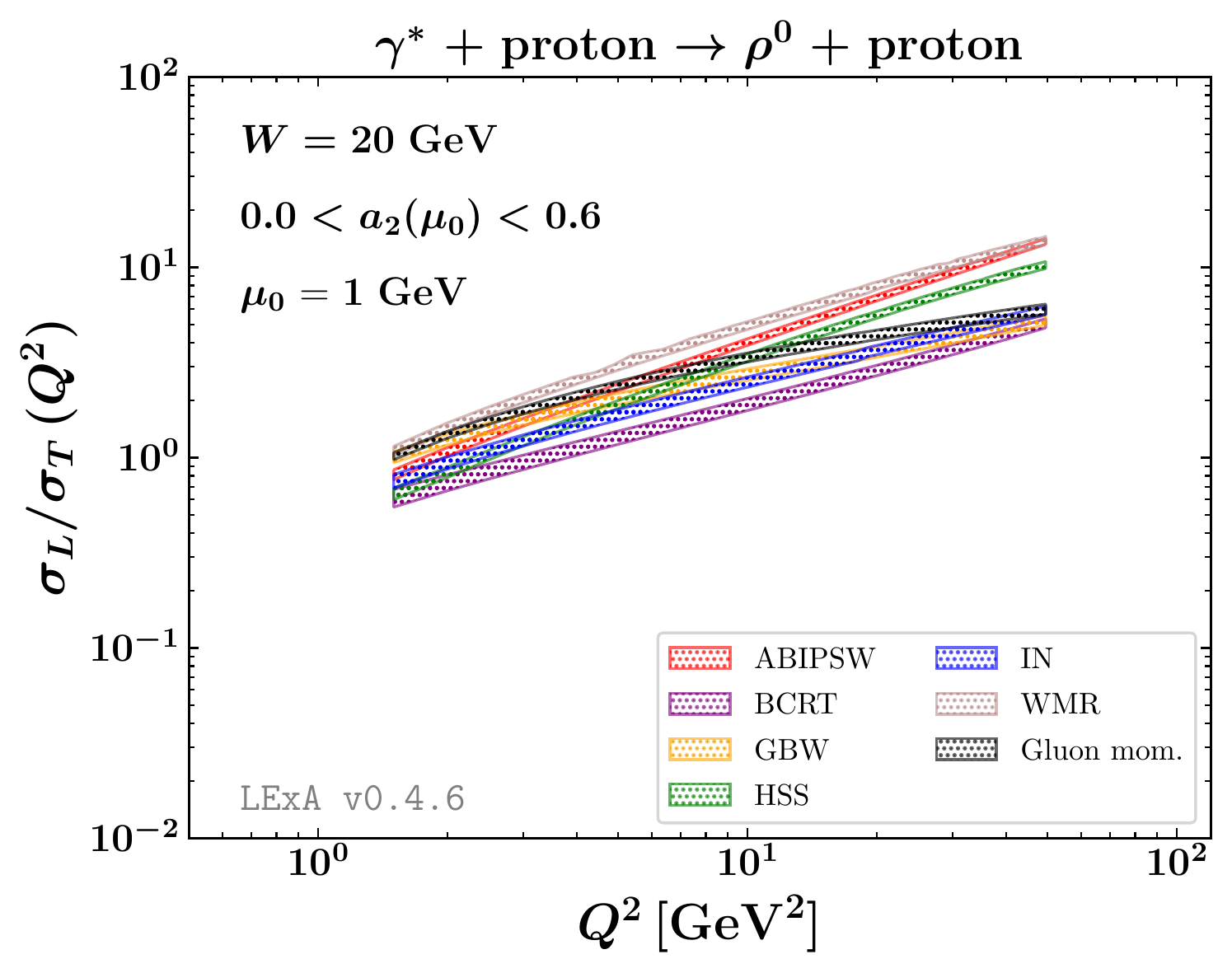} \hspace{0.50cm}
\includegraphics[width=0.38\textwidth]{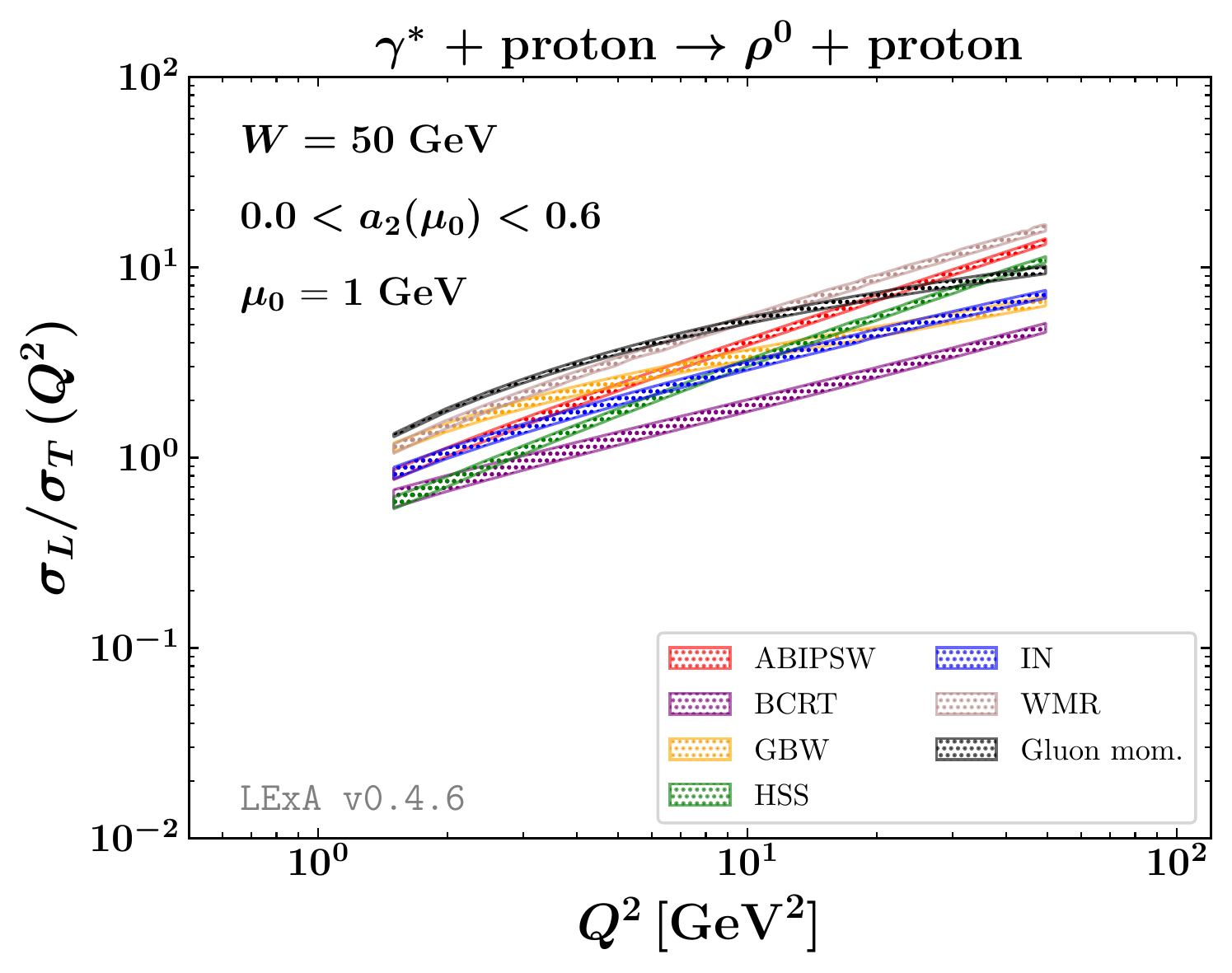}

\caption{$Q^2$-dependence polarized cross sections $\sigma_L$ (upper), $\sigma_T$ (central) and their ratio $\sigma_L/\sigma_T$ (lower), for all the considered UGD models, at $W = 20$ GeV (left) and $W = 50$ GeV (right).
Shaded bands portray the effect of varying $a_2(\mu_0 = 1\,$\rm GeV$)$ between $0.0$ and $0.6$. Figures from Ref.\tcite{Bolognino:2021niq}.}
\label{fig:sigma_LTR_all}
\end{figure*}

\section{Exclusive $\rho$-meson production at the EIC}
\label{sec:pheno}

Panels of Fig.\tref{fig:sigma_LTR_all} show the $Q^2$-behavior of predictions for the polarized cross sections $\sigma_L$ (upper), $\sigma_T$ (central), and their ratio $\sigma_L/\sigma_T$ (lower), as obtained
with all the seven UGD at two nominal values of EIC energies, $W = 20$ GeV (left) and $W = 50$ GeV (right).
Uncertainty bands are built by accounting for the va\-ria\-tion of the Gegenbauer coefficient $a_2(\mu_0)$ entering the definition of DAs (see Section~2.1 of Ref.\tcite{Bolognino:2021niq}).
We observe that the uncertainty rising from the choice of the UGD is much larger than the one associated to the variation of $a_2(\mu_0)$.
The inspection of results presented in this Section supports the statement that future data collected at the EIC will bring a high potential to constrain the UGD as well as to shed light on the gluon dynamics inside the proton at small~$x$.

\section{Summary and Outlook}
\label{sec:conclusions}

We computed cross sections for the exclusive diffractive leptoproduction of $\rho$ mesons in the energy range of forthcoming studies at the EIC. We made use of the high-energy factorization for forward helicity amplitudes~\cite{Anikin:2009bf,Anikin:2011sa}, adopting an empirical parametrization of the dif\-frac\-tion slope to get  relevant polarized cross sections.
Impact factors for forward $\rho$ me\-sons probe the transverse-momentum shape of the UGD in different ways, so that the polarization dependence of $\rho$-emission can serve as a powerful probe channel of the UGD~\cite{Bolognino:2018mlw,Bolognino:2018rhb}.
New data collected at the EIC have a potential to shed light on the intersection regime between high-energy and TMD factorization.
Future studies will extend this analysis at the next-to-leading order and will complement the information on the hadronic structure at low $x$ gathered at the EIC with the one accessible at other new-generation facilities: the Muon-Ion Collider (MuIC)\tcite{Acosta:2022ejc}, NICA\tcite{Arbuzov:2020cqg}, the Forward Physics Facility (FPF)\tcite{Anchordoqui:2021ghd,Feng:2022inv,Celiberto:2022rfj}, and the High-Luminosity LHC\tcite{Chapon:2020heu,Amoroso:2022eow}.

\section*{Acknowledgments}
\label{sec:acknowledgments}

A.D.B. and A.P. acknowledge support from the INFN/\-QFT@COL\-LI\-DERS project.
F.G.C. acknowledges support from the Italian Ministry of Education, Universities and Research under the FARE grant ``3DGLUE'' (n. R16XKPHL3N), and from the INFN/NIN\-PHA pro\-ject.
F.G.C. thanks the Universit\`a degli Studi di Pavia for the warm hospitality.
The work of D.I. was carried out within the framework of the state contract of the Sobolev Institute of Mathematics (Project No. 0314-2019-0021).
This study was partially supported by the Polish National Science Center Grant No. UMO-2018/31\-/\-B/ST2/03537 and by the Center for Innovation and Transfer of Natural Sciences and Engineering Knowledge in Rzesz\'ow.

\bibliographystyle{apsrev}
\bibliography{references}

\begin{thebibliography}{89}
\expandafter\ifx\csname natexlab\endcsname\relax\def\natexlab#1{#1}\fi
\expandafter\ifx\csname bibnamefont\endcsname\relax
  \def\bibnamefont#1{#1}\fi
\expandafter\ifx\csname bibfnamefont\endcsname\relax
  \def\bibfnamefont#1{#1}\fi
\expandafter\ifx\csname citenamefont\endcsname\relax
  \def\citenamefont#1{#1}\fi
\expandafter\ifx\csname url\endcsname\relax
  \def\url#1{\texttt{#1}}\fi
\expandafter\ifx\csname urlprefix\endcsname\relax\def\urlprefix{URL }\fi
\providecommand{\bibinfo}[2]{#2}
\providecommand{\eprint}[2][]{\url{#2}}

\bibitem[{\citenamefont{Collins and Soper}(1981)}]{Collins:1981uk}
\bibinfo{author}{\bibfnamefont{J.~C.} \bibnamefont{Collins}} \bibnamefont{and}
  \bibinfo{author}{\bibfnamefont{D.~E.} \bibnamefont{Soper}},
  \bibinfo{journal}{Nucl. Phys.} \textbf{\bibinfo{volume}{B193}},
  \bibinfo{pages}{381} (\bibinfo{year}{1981}), \bibinfo{note}{[Erratum: Nucl.
  Phys.B213,545(1983)]}.

\bibitem[{\citenamefont{Collins}(2011)}]{Collins:2011zzd}
\bibinfo{author}{\bibfnamefont{J.}~\bibnamefont{Collins}},
  \bibinfo{journal}{Camb. Monogr. Part. Phys. Nucl. Phys. Cosmol.}
  \textbf{\bibinfo{volume}{32}}, \bibinfo{pages}{1} (\bibinfo{year}{2011}).

\bibitem[{\citenamefont{Fadin et~al.}(1975)\citenamefont{Fadin, Kuraev, and
  Lipatov}}]{Fadin:1975cb}
\bibinfo{author}{\bibfnamefont{V.~S.} \bibnamefont{Fadin}},
  \bibinfo{author}{\bibfnamefont{E.}~\bibnamefont{Kuraev}}, \bibnamefont{and}
  \bibinfo{author}{\bibfnamefont{L.}~\bibnamefont{Lipatov}},
  \bibinfo{journal}{Phys. Lett. B} \textbf{\bibinfo{volume}{60}},
  \bibinfo{pages}{50} (\bibinfo{year}{1975}).

\bibitem[{\citenamefont{Kuraev et~al.}(1976)\citenamefont{Kuraev, Lipatov, and
  Fadin}}]{Kuraev:1976ge}
\bibinfo{author}{\bibfnamefont{E.~A.} \bibnamefont{Kuraev}},
  \bibinfo{author}{\bibfnamefont{L.~N.} \bibnamefont{Lipatov}},
  \bibnamefont{and} \bibinfo{author}{\bibfnamefont{V.~S.} \bibnamefont{Fadin}},
  \bibinfo{journal}{Sov. Phys. JETP} \textbf{\bibinfo{volume}{44}},
  \bibinfo{pages}{443} (\bibinfo{year}{1976}).

\bibitem[{\citenamefont{Kuraev et~al.}(1977)\citenamefont{Kuraev, Lipatov, and
  Fadin}}]{Kuraev:1977fs}
\bibinfo{author}{\bibfnamefont{E.}~\bibnamefont{Kuraev}},
  \bibinfo{author}{\bibfnamefont{L.}~\bibnamefont{Lipatov}}, \bibnamefont{and}
  \bibinfo{author}{\bibfnamefont{V.~S.} \bibnamefont{Fadin}},
  \bibinfo{journal}{Sov.\ Phys.\ JETP} \textbf{\bibinfo{volume}{45}},
  \bibinfo{pages}{199} (\bibinfo{year}{1977}).

\bibitem[{\citenamefont{Balitsky and Lipatov}(1978)}]{Balitsky:1978ic}
\bibinfo{author}{\bibfnamefont{I.}~\bibnamefont{Balitsky}} \bibnamefont{and}
  \bibinfo{author}{\bibfnamefont{L.}~\bibnamefont{Lipatov}},
  \bibinfo{journal}{Sov.\ J.\ Nucl.\ Phys.} \textbf{\bibinfo{volume}{28}},
  \bibinfo{pages}{822} (\bibinfo{year}{1978}).

\bibitem[{\citenamefont{Gribov et~al.}(1983)\citenamefont{Gribov, Levin, and
  Ryskin}}]{Gribov:1983ivg}
\bibinfo{author}{\bibfnamefont{L.~V.} \bibnamefont{Gribov}},
  \bibinfo{author}{\bibfnamefont{E.~M.} \bibnamefont{Levin}}, \bibnamefont{and}
  \bibinfo{author}{\bibfnamefont{M.~G.} \bibnamefont{Ryskin}},
  \bibinfo{journal}{Phys. Rept.} \textbf{\bibinfo{volume}{100}},
  \bibinfo{pages}{1} (\bibinfo{year}{1983}).

\bibitem[{\citenamefont{Celiberto}(2017)}]{Celiberto:2017ius}
\bibinfo{author}{\bibfnamefont{F.~G.} \bibnamefont{Celiberto}}, Ph.D. thesis,
  \bibinfo{school}{Universit\`a della Calabria and INFN-Cosenza}
  (\bibinfo{year}{2017}), \eprint{1707.04315}.

\bibitem[{\citenamefont{Colferai et~al.}(2010)\citenamefont{Colferai,
  Schwennsen, Szymanowski, and Wallon}}]{Colferai:2010wu}
\bibinfo{author}{\bibfnamefont{D.}~\bibnamefont{Colferai}},
  \bibinfo{author}{\bibfnamefont{F.}~\bibnamefont{Schwennsen}},
  \bibinfo{author}{\bibfnamefont{L.}~\bibnamefont{Szymanowski}},
  \bibnamefont{and} \bibinfo{author}{\bibfnamefont{S.}~\bibnamefont{Wallon}},
  \bibinfo{journal}{JHEP} \textbf{\bibinfo{volume}{12}}, \bibinfo{pages}{026}
  (\bibinfo{year}{2010}), \eprint{1002.1365}.

\bibitem[{\citenamefont{Deak et~al.}(2009)\citenamefont{Deak, Hautmann, Jung,
  and Kutak}}]{Deak:2009xt}
\bibinfo{author}{\bibfnamefont{M.}~\bibnamefont{Deak}},
  \bibinfo{author}{\bibfnamefont{F.}~\bibnamefont{Hautmann}},
  \bibinfo{author}{\bibfnamefont{H.}~\bibnamefont{Jung}}, \bibnamefont{and}
  \bibinfo{author}{\bibfnamefont{K.}~\bibnamefont{Kutak}},
  \bibinfo{journal}{JHEP} \textbf{\bibinfo{volume}{09}}, \bibinfo{pages}{121}
  (\bibinfo{year}{2009}), \eprint{0908.0538}.

\bibitem[{\citenamefont{Deak et~al.}(2019)\citenamefont{Deak, van Hameren,
  Jung, Kusina, Kutak, and Serino}}]{Deak:2018obv}
\bibinfo{author}{\bibfnamefont{M.}~\bibnamefont{Deak}},
  \bibinfo{author}{\bibfnamefont{A.}~\bibnamefont{van Hameren}},
  \bibinfo{author}{\bibfnamefont{H.}~\bibnamefont{Jung}},
  \bibinfo{author}{\bibfnamefont{A.}~\bibnamefont{Kusina}},
  \bibinfo{author}{\bibfnamefont{K.}~\bibnamefont{Kutak}}, \bibnamefont{and}
  \bibinfo{author}{\bibfnamefont{M.}~\bibnamefont{Serino}},
  \bibinfo{journal}{Phys. Rev. D} \textbf{\bibinfo{volume}{99}},
  \bibinfo{pages}{094011} (\bibinfo{year}{2019}), \eprint{1809.03854}.

\bibitem[{\citenamefont{Blanco et~al.}(2020)\citenamefont{Blanco, van Hameren,
  Kotko, and Kutak}}]{Blanco:2020akb}
\bibinfo{author}{\bibfnamefont{E.}~\bibnamefont{Blanco}},
  \bibinfo{author}{\bibfnamefont{A.}~\bibnamefont{van Hameren}},
  \bibinfo{author}{\bibfnamefont{P.}~\bibnamefont{Kotko}}, \bibnamefont{and}
  \bibinfo{author}{\bibfnamefont{K.}~\bibnamefont{Kutak}},
  \bibinfo{journal}{JHEP} \textbf{\bibinfo{volume}{12}}, \bibinfo{pages}{158}
  (\bibinfo{year}{2020}), \eprint{2008.07916}.

\bibitem[{\citenamefont{van Hameren et~al.}(2021)\citenamefont{van Hameren,
  Kotko, Kutak, and Sapeta}}]{vanHameren:2020rqt}
\bibinfo{author}{\bibfnamefont{A.}~\bibnamefont{van Hameren}},
  \bibinfo{author}{\bibfnamefont{P.}~\bibnamefont{Kotko}},
  \bibinfo{author}{\bibfnamefont{K.}~\bibnamefont{Kutak}}, \bibnamefont{and}
  \bibinfo{author}{\bibfnamefont{S.}~\bibnamefont{Sapeta}},
  \bibinfo{journal}{Phys. Lett. B} \textbf{\bibinfo{volume}{814}},
  \bibinfo{pages}{136078} (\bibinfo{year}{2021}), \eprint{2010.13066}.

\bibitem[{\citenamefont{van Hameren et~al.}(2022)\citenamefont{van Hameren,
  Motyka, and Ziarko}}]{vanHameren:2022mtk}
\bibinfo{author}{\bibfnamefont{A.}~\bibnamefont{van Hameren}},
  \bibinfo{author}{\bibfnamefont{L.}~\bibnamefont{Motyka}}, \bibnamefont{and}
  \bibinfo{author}{\bibfnamefont{G.}~\bibnamefont{Ziarko}}
  (\bibinfo{year}{2022}), \eprint{2205.09585}.

\bibitem[{\citenamefont{Duclou\'e et~al.}(2014)\citenamefont{Duclou\'e,
  Szymanowski, and Wallon}}]{Ducloue:2013bva}
\bibinfo{author}{\bibfnamefont{B.}~\bibnamefont{Duclou\'e}},
  \bibinfo{author}{\bibfnamefont{L.}~\bibnamefont{Szymanowski}},
  \bibnamefont{and} \bibinfo{author}{\bibfnamefont{S.}~\bibnamefont{Wallon}},
  \bibinfo{journal}{Phys. Rev. Lett.} \textbf{\bibinfo{volume}{112}},
  \bibinfo{pages}{082003} (\bibinfo{year}{2014}), \eprint{1309.3229}.

\bibitem[{\citenamefont{Caporale et~al.}(2014)\citenamefont{Caporale, Ivanov,
  Murdaca, and Papa}}]{Caporale:2014gpa}
\bibinfo{author}{\bibfnamefont{F.}~\bibnamefont{Caporale}},
  \bibinfo{author}{\bibfnamefont{D.~{\relax Yu}.} \bibnamefont{Ivanov}},
  \bibinfo{author}{\bibfnamefont{B.}~\bibnamefont{Murdaca}}, \bibnamefont{and}
  \bibinfo{author}{\bibfnamefont{A.}~\bibnamefont{Papa}},
  \bibinfo{journal}{Eur. Phys. J. C} \textbf{\bibinfo{volume}{74}},
  \bibinfo{pages}{3084} (\bibinfo{year}{2014}), \bibinfo{note}{[Erratum:
  Eur.Phys.J.C 75, 535 (2015)]}, \eprint{1407.8431}.

\bibitem[{\citenamefont{Celiberto
  et~al.}(2015{\natexlab{a}})\citenamefont{Celiberto, Ivanov, Murdaca, and
  Papa}}]{Celiberto:2015yba}
\bibinfo{author}{\bibfnamefont{F.~G.} \bibnamefont{Celiberto}},
  \bibinfo{author}{\bibfnamefont{D.~{\relax Yu}.} \bibnamefont{Ivanov}},
  \bibinfo{author}{\bibfnamefont{B.}~\bibnamefont{Murdaca}}, \bibnamefont{and}
  \bibinfo{author}{\bibfnamefont{A.}~\bibnamefont{Papa}},
  \bibinfo{journal}{Eur. Phys. J. C} \textbf{\bibinfo{volume}{75}},
  \bibinfo{pages}{292} (\bibinfo{year}{2015}{\natexlab{a}}),
  \eprint{1504.08233}.

\bibitem[{\citenamefont{Celiberto
  et~al.}(2015{\natexlab{b}})\citenamefont{Celiberto, Ivanov, Murdaca, and
  Papa}}]{Celiberto:2015mpa}
\bibinfo{author}{\bibfnamefont{F.~G.} \bibnamefont{Celiberto}},
  \bibinfo{author}{\bibfnamefont{D.~{\relax Yu}.} \bibnamefont{Ivanov}},
  \bibinfo{author}{\bibfnamefont{B.}~\bibnamefont{Murdaca}}, \bibnamefont{and}
  \bibinfo{author}{\bibfnamefont{A.}~\bibnamefont{Papa}},
  \bibinfo{journal}{Acta Phys. Polon. Supp.} \textbf{\bibinfo{volume}{8}},
  \bibinfo{pages}{935} (\bibinfo{year}{2015}{\natexlab{b}}),
  \eprint{1510.01626}.

\bibitem[{\citenamefont{Celiberto et~al.}(2016)\citenamefont{Celiberto, Ivanov,
  Murdaca, and Papa}}]{Celiberto:2016ygs}
\bibinfo{author}{\bibfnamefont{F.~G.} \bibnamefont{Celiberto}},
  \bibinfo{author}{\bibfnamefont{D.~{\relax Yu}.} \bibnamefont{Ivanov}},
  \bibinfo{author}{\bibfnamefont{B.}~\bibnamefont{Murdaca}}, \bibnamefont{and}
  \bibinfo{author}{\bibfnamefont{A.}~\bibnamefont{Papa}},
  \bibinfo{journal}{Eur. Phys. J. C} \textbf{\bibinfo{volume}{76}},
  \bibinfo{pages}{224} (\bibinfo{year}{2016}), \eprint{1601.07847}.

\bibitem[{\citenamefont{Celiberto}(2016)}]{Celiberto:2016vhn}
\bibinfo{author}{\bibfnamefont{F.~G.} \bibnamefont{Celiberto}},
  \bibinfo{journal}{Frascati Phys. Ser.} \textbf{\bibinfo{volume}{63}},
  \bibinfo{pages}{43} (\bibinfo{year}{2016}), \eprint{1606.07327}.

\bibitem[{\citenamefont{Caporale et~al.}(2018)\citenamefont{Caporale,
  Celiberto, Chachamis, Gordo~G\'omez, and Sabio~Vera}}]{Caporale:2018qnm}
\bibinfo{author}{\bibfnamefont{F.}~\bibnamefont{Caporale}},
  \bibinfo{author}{\bibfnamefont{F.~G.} \bibnamefont{Celiberto}},
  \bibinfo{author}{\bibfnamefont{G.}~\bibnamefont{Chachamis}},
  \bibinfo{author}{\bibfnamefont{D.}~\bibnamefont{Gordo~G\'omez}},
  \bibnamefont{and}
  \bibinfo{author}{\bibfnamefont{A.}~\bibnamefont{Sabio~Vera}},
  \bibinfo{journal}{Nucl. Phys. B} \textbf{\bibinfo{volume}{935}},
  \bibinfo{pages}{412} (\bibinfo{year}{2018}), \eprint{1806.06309}.

\bibitem[{\citenamefont{Celiberto and Papa}(2022)}]{Celiberto:2022gji}
\bibinfo{author}{\bibfnamefont{F.~G.} \bibnamefont{Celiberto}}
  \bibnamefont{and} \bibinfo{author}{\bibfnamefont{A.}~\bibnamefont{Papa}}
  (\bibinfo{year}{2022}), \eprint{2207.05015}.

\bibitem[{\citenamefont{Celiberto
  et~al.}(2018{\natexlab{a}})\citenamefont{Celiberto, Ivanov, Murdaca, and
  Papa}}]{Celiberto:2017nyx}
\bibinfo{author}{\bibfnamefont{F.~G.} \bibnamefont{Celiberto}},
  \bibinfo{author}{\bibfnamefont{D.~{\relax Yu}.} \bibnamefont{Ivanov}},
  \bibinfo{author}{\bibfnamefont{B.}~\bibnamefont{Murdaca}}, \bibnamefont{and}
  \bibinfo{author}{\bibfnamefont{A.}~\bibnamefont{Papa}},
  \bibinfo{journal}{Phys. Lett. B} \textbf{\bibinfo{volume}{777}},
  \bibinfo{pages}{141} (\bibinfo{year}{2018}{\natexlab{a}}),
  \eprint{1709.10032}.

\bibitem[{\citenamefont{Bolognino
  et~al.}(2019{\natexlab{a}})\citenamefont{Bolognino, Celiberto, Fucilla,
  Ivanov, Murdaca, and Papa}}]{Bolognino:2019ccd}
\bibinfo{author}{\bibfnamefont{A.~D.} \bibnamefont{Bolognino}},
  \bibinfo{author}{\bibfnamefont{F.~G.} \bibnamefont{Celiberto}},
  \bibinfo{author}{\bibfnamefont{M.}~\bibnamefont{Fucilla}},
  \bibinfo{author}{\bibfnamefont{D.~{\relax Yu}.} \bibnamefont{Ivanov}},
  \bibinfo{author}{\bibfnamefont{B.}~\bibnamefont{Murdaca}}, \bibnamefont{and}
  \bibinfo{author}{\bibfnamefont{A.}~\bibnamefont{Papa}},
  \bibinfo{journal}{PoS} \textbf{\bibinfo{volume}{DIS2019}},
  \bibinfo{pages}{067} (\bibinfo{year}{2019}{\natexlab{a}}),
  \eprint{1906.05940}.

\bibitem[{\citenamefont{Bolognino
  et~al.}(2019{\natexlab{b}})\citenamefont{Bolognino, Celiberto, Fucilla,
  Ivanov, and Papa}}]{Bolognino:2019yls}
\bibinfo{author}{\bibfnamefont{A.~D.} \bibnamefont{Bolognino}},
  \bibinfo{author}{\bibfnamefont{F.~G.} \bibnamefont{Celiberto}},
  \bibinfo{author}{\bibfnamefont{M.}~\bibnamefont{Fucilla}},
  \bibinfo{author}{\bibfnamefont{D.~{\relax Yu}.} \bibnamefont{Ivanov}},
  \bibnamefont{and} \bibinfo{author}{\bibfnamefont{A.}~\bibnamefont{Papa}},
  \bibinfo{journal}{Eur. Phys. J. C} \textbf{\bibinfo{volume}{79}},
  \bibinfo{pages}{939} (\bibinfo{year}{2019}{\natexlab{b}}),
  \eprint{1909.03068}.

\bibitem[{\citenamefont{Bolognino
  et~al.}(2021{\natexlab{a}})\citenamefont{Bolognino, Celiberto, Fucilla,
  Ivanov, and Papa}}]{Bolognino:2021mrc}
\bibinfo{author}{\bibfnamefont{A.~D.} \bibnamefont{Bolognino}},
  \bibinfo{author}{\bibfnamefont{F.~G.} \bibnamefont{Celiberto}},
  \bibinfo{author}{\bibfnamefont{M.}~\bibnamefont{Fucilla}},
  \bibinfo{author}{\bibfnamefont{D.~{\relax Yu}.} \bibnamefont{Ivanov}},
  \bibnamefont{and} \bibinfo{author}{\bibfnamefont{A.}~\bibnamefont{Papa}},
  \bibinfo{journal}{Phys. Rev. D} \textbf{\bibinfo{volume}{103}},
  \bibinfo{pages}{094004} (\bibinfo{year}{2021}{\natexlab{a}}),
  \eprint{2103.07396}.

\bibitem[{\citenamefont{Maciula et~al.}(2022)\citenamefont{Maciula, Pasechnik,
  and Szczurek}}]{Maciula:2022bfv}
\bibinfo{author}{\bibfnamefont{R.}~\bibnamefont{Maciula}},
  \bibinfo{author}{\bibfnamefont{R.}~\bibnamefont{Pasechnik}},
  \bibnamefont{and} \bibinfo{author}{\bibfnamefont{A.}~\bibnamefont{Szczurek}}
  (\bibinfo{year}{2022}), \eprint{2202.07585}.

\bibitem[{\citenamefont{Bolognino
  et~al.}(2018{\natexlab{a}})\citenamefont{Bolognino, Celiberto, Ivanov,
  Mohammed, and Papa}}]{Bolognino:2018oth}
\bibinfo{author}{\bibfnamefont{A.~D.} \bibnamefont{Bolognino}},
  \bibinfo{author}{\bibfnamefont{F.~G.} \bibnamefont{Celiberto}},
  \bibinfo{author}{\bibfnamefont{D.~{\relax Yu}.} \bibnamefont{Ivanov}},
  \bibinfo{author}{\bibfnamefont{M.~M.} \bibnamefont{Mohammed}},
  \bibnamefont{and} \bibinfo{author}{\bibfnamefont{A.}~\bibnamefont{Papa}},
  \bibinfo{journal}{Eur. Phys. J. C} \textbf{\bibinfo{volume}{78}},
  \bibinfo{pages}{772} (\bibinfo{year}{2018}{\natexlab{a}}),
  \eprint{1808.05483}.

\bibitem[{\citenamefont{Bolognino
  et~al.}(2019{\natexlab{c}})\citenamefont{Bolognino, Celiberto, Ivanov,
  Mohammed, and Papa}}]{Bolognino:2019yqj}
\bibinfo{author}{\bibfnamefont{A.~D.} \bibnamefont{Bolognino}},
  \bibinfo{author}{\bibfnamefont{F.~G.} \bibnamefont{Celiberto}},
  \bibinfo{author}{\bibfnamefont{D.~{\relax Yu}.} \bibnamefont{Ivanov}},
  \bibinfo{author}{\bibfnamefont{M.~M.} \bibnamefont{Mohammed}},
  \bibnamefont{and} \bibinfo{author}{\bibfnamefont{A.}~\bibnamefont{Papa}},
  \bibinfo{journal}{Acta Phys. Polon. Supp.} \textbf{\bibinfo{volume}{12}},
  \bibinfo{pages}{773} (\bibinfo{year}{2019}{\natexlab{c}}),
  \eprint{1902.04511}.

\bibitem[{\citenamefont{Celiberto et~al.}(2020)\citenamefont{Celiberto, Ivanov,
  and Papa}}]{Celiberto:2020rxb}
\bibinfo{author}{\bibfnamefont{F.~G.} \bibnamefont{Celiberto}},
  \bibinfo{author}{\bibfnamefont{D.~{\relax Yu}.} \bibnamefont{Ivanov}},
  \bibnamefont{and} \bibinfo{author}{\bibfnamefont{A.}~\bibnamefont{Papa}},
  \bibinfo{journal}{Phys. Rev. D} \textbf{\bibinfo{volume}{102}},
  \bibinfo{pages}{094019} (\bibinfo{year}{2020}), \eprint{2008.10513}.

\bibitem[{\citenamefont{Boussarie et~al.}(2018)\citenamefont{Boussarie,
  Duclou\'e, Szymanowski, and Wallon}}]{Boussarie:2017oae}
\bibinfo{author}{\bibfnamefont{R.}~\bibnamefont{Boussarie}},
  \bibinfo{author}{\bibfnamefont{B.}~\bibnamefont{Duclou\'e}},
  \bibinfo{author}{\bibfnamefont{L.}~\bibnamefont{Szymanowski}},
  \bibnamefont{and} \bibinfo{author}{\bibfnamefont{S.}~\bibnamefont{Wallon}},
  \bibinfo{journal}{Phys. Rev. D} \textbf{\bibinfo{volume}{97}},
  \bibinfo{pages}{014008} (\bibinfo{year}{2018}), \eprint{1709.01380}.

\bibitem[{\citenamefont{Celiberto
  et~al.}(2021{\natexlab{a}})\citenamefont{Celiberto, Fucilla, Ivanov, and
  Papa}}]{Celiberto:2021dzy}
\bibinfo{author}{\bibfnamefont{F.~G.} \bibnamefont{Celiberto}},
  \bibinfo{author}{\bibfnamefont{M.}~\bibnamefont{Fucilla}},
  \bibinfo{author}{\bibfnamefont{D.~{\relax Yu}.} \bibnamefont{Ivanov}},
  \bibnamefont{and} \bibinfo{author}{\bibfnamefont{A.}~\bibnamefont{Papa}},
  \bibinfo{journal}{Eur. Phys. J. C} \textbf{\bibinfo{volume}{81}},
  \bibinfo{pages}{780} (\bibinfo{year}{2021}{\natexlab{a}}),
  \eprint{2105.06432}.

\bibitem[{\citenamefont{Guiot and van Hameren}(2021)}]{Guiot:2021vnp}
\bibinfo{author}{\bibfnamefont{B.}~\bibnamefont{Guiot}} \bibnamefont{and}
  \bibinfo{author}{\bibfnamefont{A.}~\bibnamefont{van Hameren}},
  \bibinfo{journal}{Phys. Rev. D} \textbf{\bibinfo{volume}{104}},
  \bibinfo{pages}{094038} (\bibinfo{year}{2021}), \eprint{2108.06419}.

\bibitem[{\citenamefont{Celiberto
  et~al.}(2021{\natexlab{b}})\citenamefont{Celiberto, Fucilla, Ivanov,
  Mohammed, and Papa}}]{Celiberto:2021fdp}
\bibinfo{author}{\bibfnamefont{F.~G.} \bibnamefont{Celiberto}},
  \bibinfo{author}{\bibfnamefont{M.}~\bibnamefont{Fucilla}},
  \bibinfo{author}{\bibfnamefont{D.~{\relax Yu}.} \bibnamefont{Ivanov}},
  \bibinfo{author}{\bibfnamefont{M.~M.~A.} \bibnamefont{Mohammed}},
  \bibnamefont{and} \bibinfo{author}{\bibfnamefont{A.}~\bibnamefont{Papa}},
  \bibinfo{journal}{Phys. Rev. D} \textbf{\bibinfo{volume}{104}},
  \bibinfo{pages}{114007} (\bibinfo{year}{2021}{\natexlab{b}}),
  \eprint{2109.11875}.

\bibitem[{\citenamefont{Celiberto and
  Fucilla}(2022{\natexlab{a}})}]{Celiberto:2022dyf}
\bibinfo{author}{\bibfnamefont{F.~G.} \bibnamefont{Celiberto}}
  \bibnamefont{and} \bibinfo{author}{\bibfnamefont{M.}~\bibnamefont{Fucilla}},
  \bibinfo{journal}{under revision in Eur. Phys. J. C}
  (\bibinfo{year}{2022}{\natexlab{a}}), \eprint{2202.12227}.

\bibitem[{\citenamefont{Celiberto}(2022{\natexlab{a}})}]{Celiberto:2022keu}
\bibinfo{author}{\bibfnamefont{F.~G.} \bibnamefont{Celiberto}}
  (\bibinfo{year}{2022}{\natexlab{a}}), \eprint{2206.09413}.

\bibitem[{\citenamefont{Celiberto and
  Fucilla}(2022{\natexlab{b}})}]{Celiberto:2022kza}
\bibinfo{author}{\bibfnamefont{F.~G.} \bibnamefont{Celiberto}}
  \bibnamefont{and} \bibinfo{author}{\bibfnamefont{M.}~\bibnamefont{Fucilla}},
  \bibinfo{journal}{Zenodo, in press}  (\bibinfo{year}{2022}{\natexlab{b}}),
  \eprint{2208.07206}.

\bibitem[{\citenamefont{Celiberto
  et~al.}(2021{\natexlab{c}})\citenamefont{Celiberto, Ivanov, Mohammed, and
  Papa}}]{Celiberto:2020tmb}
\bibinfo{author}{\bibfnamefont{F.~G.} \bibnamefont{Celiberto}},
  \bibinfo{author}{\bibfnamefont{D.~{\relax Yu}.} \bibnamefont{Ivanov}},
  \bibinfo{author}{\bibfnamefont{M.~M.~A.} \bibnamefont{Mohammed}},
  \bibnamefont{and} \bibinfo{author}{\bibfnamefont{A.}~\bibnamefont{Papa}},
  \bibinfo{journal}{Eur. Phys. J. C} \textbf{\bibinfo{volume}{81}},
  \bibinfo{pages}{293} (\bibinfo{year}{2021}{\natexlab{c}}),
  \eprint{2008.00501}.

\bibitem[{\citenamefont{Celiberto
  et~al.}(2022{\natexlab{a}})\citenamefont{Celiberto, Fucilla, Mohammed, and
  Papa}}]{Celiberto:2022zdg}
\bibinfo{author}{\bibfnamefont{F.~G.} \bibnamefont{Celiberto}},
  \bibinfo{author}{\bibfnamefont{M.}~\bibnamefont{Fucilla}},
  \bibinfo{author}{\bibfnamefont{M.~M.~A.} \bibnamefont{Mohammed}},
  \bibnamefont{and} \bibinfo{author}{\bibfnamefont{A.}~\bibnamefont{Papa}},
  \bibinfo{journal}{Phys. Rev. D} \textbf{\bibinfo{volume}{105}},
  \bibinfo{pages}{114056} (\bibinfo{year}{2022}{\natexlab{a}}),
  \eprint{2205.13429}.

\bibitem[{\citenamefont{Hentschinski et~al.}(2021)\citenamefont{Hentschinski,
  Kutak, and van Hameren}}]{Hentschinski:2020tbi}
\bibinfo{author}{\bibfnamefont{M.}~\bibnamefont{Hentschinski}},
  \bibinfo{author}{\bibfnamefont{K.}~\bibnamefont{Kutak}}, \bibnamefont{and}
  \bibinfo{author}{\bibfnamefont{A.}~\bibnamefont{van Hameren}},
  \bibinfo{journal}{Eur. Phys. J. C} \textbf{\bibinfo{volume}{81}},
  \bibinfo{pages}{112} (\bibinfo{year}{2021}), \bibinfo{note}{[Erratum: Eur.
  Phys. J. C 81, 262 (2021)]}, \eprint{2011.03193}.

\bibitem[{\citenamefont{Celiberto
  et~al.}(2022{\natexlab{b}})\citenamefont{Celiberto, Fucilla, Ivanov,
  Mohammed, and Papa}}]{Celiberto:2022fgx}
\bibinfo{author}{\bibfnamefont{F.~G.} \bibnamefont{Celiberto}},
  \bibinfo{author}{\bibfnamefont{M.}~\bibnamefont{Fucilla}},
  \bibinfo{author}{\bibfnamefont{D.~Y.} \bibnamefont{Ivanov}},
  \bibinfo{author}{\bibfnamefont{M.~M.~A.} \bibnamefont{Mohammed}},
  \bibnamefont{and} \bibinfo{author}{\bibfnamefont{A.}~\bibnamefont{Papa}},
  \bibinfo{journal}{JHEP} \textbf{\bibinfo{volume}{08}}, \bibinfo{pages}{092}
  (\bibinfo{year}{2022}{\natexlab{b}}), \eprint{2205.02681}.

\bibitem[{\citenamefont{Motyka et~al.}(2015)\citenamefont{Motyka, Sadzikowski,
  and Stebel}}]{Motyka:2014lya}
\bibinfo{author}{\bibfnamefont{L.}~\bibnamefont{Motyka}},
  \bibinfo{author}{\bibfnamefont{M.}~\bibnamefont{Sadzikowski}},
  \bibnamefont{and} \bibinfo{author}{\bibfnamefont{T.}~\bibnamefont{Stebel}},
  \bibinfo{journal}{JHEP} \textbf{\bibinfo{volume}{05}}, \bibinfo{pages}{087}
  (\bibinfo{year}{2015}), \eprint{1412.4675}.

\bibitem[{\citenamefont{Brzeminski et~al.}(2017)\citenamefont{Brzeminski,
  Motyka, Sadzikowski, and Stebel}}]{Brzeminski:2016lwh}
\bibinfo{author}{\bibfnamefont{D.}~\bibnamefont{Brzeminski}},
  \bibinfo{author}{\bibfnamefont{L.}~\bibnamefont{Motyka}},
  \bibinfo{author}{\bibfnamefont{M.}~\bibnamefont{Sadzikowski}},
  \bibnamefont{and} \bibinfo{author}{\bibfnamefont{T.}~\bibnamefont{Stebel}},
  \bibinfo{journal}{JHEP} \textbf{\bibinfo{volume}{01}}, \bibinfo{pages}{005}
  (\bibinfo{year}{2017}), \eprint{1611.04449}.

\bibitem[{\citenamefont{Motyka et~al.}(2017)\citenamefont{Motyka, Sadzikowski,
  and Stebel}}]{Motyka:2016lta}
\bibinfo{author}{\bibfnamefont{L.}~\bibnamefont{Motyka}},
  \bibinfo{author}{\bibfnamefont{M.}~\bibnamefont{Sadzikowski}},
  \bibnamefont{and} \bibinfo{author}{\bibfnamefont{T.}~\bibnamefont{Stebel}},
  \bibinfo{journal}{Phys. Rev.} \textbf{\bibinfo{volume}{D95}},
  \bibinfo{pages}{114025} (\bibinfo{year}{2017}), \eprint{1609.04300}.

\bibitem[{\citenamefont{Celiberto
  et~al.}(2018{\natexlab{b}})\citenamefont{Celiberto, Gordo~G\'omez, and
  Sabio~Vera}}]{Celiberto:2018muu}
\bibinfo{author}{\bibfnamefont{F.~G.} \bibnamefont{Celiberto}},
  \bibinfo{author}{\bibfnamefont{D.}~\bibnamefont{Gordo~G\'omez}},
  \bibnamefont{and}
  \bibinfo{author}{\bibfnamefont{A.}~\bibnamefont{Sabio~Vera}},
  \bibinfo{journal}{Phys. Lett.} \textbf{\bibinfo{volume}{B786}},
  \bibinfo{pages}{201} (\bibinfo{year}{2018}{\natexlab{b}}),
  \eprint{1808.09511}.

\bibitem[{\citenamefont{Golec-Biernat et~al.}(2018)\citenamefont{Golec-Biernat,
  Motyka, and Stebel}}]{Golec-Biernat:2018kem}
\bibinfo{author}{\bibfnamefont{K.}~\bibnamefont{Golec-Biernat}},
  \bibinfo{author}{\bibfnamefont{L.}~\bibnamefont{Motyka}}, \bibnamefont{and}
  \bibinfo{author}{\bibfnamefont{T.}~\bibnamefont{Stebel}},
  \bibinfo{journal}{JHEP} \textbf{\bibinfo{volume}{12}}, \bibinfo{pages}{091}
  (\bibinfo{year}{2018}), \eprint{1811.04361}.

\bibitem[{\citenamefont{Ball et~al.}(2018)\citenamefont{Ball, Bertone, Bonvini,
  Marzani, Rojo, and Rottoli}}]{Ball:2017otu}
\bibinfo{author}{\bibfnamefont{R.~D.} \bibnamefont{Ball}},
  \bibinfo{author}{\bibfnamefont{V.}~\bibnamefont{Bertone}},
  \bibinfo{author}{\bibfnamefont{M.}~\bibnamefont{Bonvini}},
  \bibinfo{author}{\bibfnamefont{S.}~\bibnamefont{Marzani}},
  \bibinfo{author}{\bibfnamefont{J.}~\bibnamefont{Rojo}}, \bibnamefont{and}
  \bibinfo{author}{\bibfnamefont{L.}~\bibnamefont{Rottoli}},
  \bibinfo{journal}{Eur. Phys. J.} \textbf{\bibinfo{volume}{C78}},
  \bibinfo{pages}{321} (\bibinfo{year}{2018}), \eprint{1710.05935}.

\bibitem[{\citenamefont{Bonvini and Giuli}(2019)}]{Bonvini:2019wxf}
\bibinfo{author}{\bibfnamefont{M.}~\bibnamefont{Bonvini}} \bibnamefont{and}
  \bibinfo{author}{\bibfnamefont{F.}~\bibnamefont{Giuli}},
  \bibinfo{journal}{Eur. Phys. J. Plus} \textbf{\bibinfo{volume}{134}},
  \bibinfo{pages}{531} (\bibinfo{year}{2019}), \eprint{1902.11125}.

\bibitem[{\citenamefont{Bacchetta et~al.}(2020)\citenamefont{Bacchetta,
  Celiberto, Radici, and Taels}}]{Bacchetta:2020vty}
\bibinfo{author}{\bibfnamefont{A.}~\bibnamefont{Bacchetta}},
  \bibinfo{author}{\bibfnamefont{F.~G.} \bibnamefont{Celiberto}},
  \bibinfo{author}{\bibfnamefont{M.}~\bibnamefont{Radici}}, \bibnamefont{and}
  \bibinfo{author}{\bibfnamefont{P.}~\bibnamefont{Taels}},
  \bibinfo{journal}{Eur. Phys. J. C} \textbf{\bibinfo{volume}{80}},
  \bibinfo{pages}{733} (\bibinfo{year}{2020}), \eprint{2005.02288}.

\bibitem[{\citenamefont{Celiberto}(2021)}]{Celiberto:2021zww}
\bibinfo{author}{\bibfnamefont{F.~G.} \bibnamefont{Celiberto}},
  \bibinfo{journal}{Nuovo Cim.} \textbf{\bibinfo{volume}{C44}},
  \bibinfo{pages}{36} (\bibinfo{year}{2021}), \eprint{2101.04630}.

\bibitem[{\citenamefont{Bacchetta et~al.}(2022)\citenamefont{Bacchetta,
  Celiberto, Radici, and Signori}}]{Bacchetta:2022nyv}
\bibinfo{author}{\bibfnamefont{A.}~\bibnamefont{Bacchetta}},
  \bibinfo{author}{\bibfnamefont{F.~G.} \bibnamefont{Celiberto}},
  \bibinfo{author}{\bibfnamefont{M.}~\bibnamefont{Radici}}, \bibnamefont{and}
  \bibinfo{author}{\bibfnamefont{A.}~\bibnamefont{Signori}},
  \bibinfo{journal}{Zenodo, in press}  (\bibinfo{year}{2022}),
  \eprint{2208.06252}.

\bibitem[{\citenamefont{Hentschinski
  et~al.}(2013{\natexlab{a}})\citenamefont{Hentschinski, Sabio~Vera, and
  Salas}}]{Hentschinski:2012kr}
\bibinfo{author}{\bibfnamefont{M.}~\bibnamefont{Hentschinski}},
  \bibinfo{author}{\bibfnamefont{A.}~\bibnamefont{Sabio~Vera}},
  \bibnamefont{and} \bibinfo{author}{\bibfnamefont{C.}~\bibnamefont{Salas}},
  \bibinfo{journal}{Phys. Rev. Lett.} \textbf{\bibinfo{volume}{110}},
  \bibinfo{pages}{041601} (\bibinfo{year}{2013}{\natexlab{a}}),
  \eprint{1209.1353}.

\bibitem[{\citenamefont{Hentschinski
  et~al.}(2013{\natexlab{b}})\citenamefont{Hentschinski, Sabio~Vera, and
  Salas}}]{Hentschinski:2013id}
\bibinfo{author}{\bibfnamefont{M.}~\bibnamefont{Hentschinski}},
  \bibinfo{author}{\bibfnamefont{A.}~\bibnamefont{Sabio~Vera}},
  \bibnamefont{and} \bibinfo{author}{\bibfnamefont{C.}~\bibnamefont{Salas}},
  \bibinfo{journal}{Phys. Rev. D} \textbf{\bibinfo{volume}{87}},
  \bibinfo{pages}{076005} (\bibinfo{year}{2013}{\natexlab{b}}),
  \eprint{1301.5283}.

\bibitem[{\citenamefont{Anikin et~al.}(2010)\citenamefont{Anikin, Ivanov, Pire,
  Szymanowski, and Wallon}}]{Anikin:2009bf}
\bibinfo{author}{\bibfnamefont{I.}~\bibnamefont{Anikin}},
  \bibinfo{author}{\bibfnamefont{D.~{\relax Yu}.} \bibnamefont{Ivanov}},
  \bibinfo{author}{\bibfnamefont{B.}~\bibnamefont{Pire}},
  \bibinfo{author}{\bibfnamefont{L.}~\bibnamefont{Szymanowski}},
  \bibnamefont{and} \bibinfo{author}{\bibfnamefont{S.}~\bibnamefont{Wallon}},
  \bibinfo{journal}{Nucl. Phys. B} \textbf{\bibinfo{volume}{828}},
  \bibinfo{pages}{1} (\bibinfo{year}{2010}), \eprint{0909.4090}.

\bibitem[{\citenamefont{Anikin et~al.}(2011)\citenamefont{Anikin, Besse,
  Ivanov, Pire, Szymanowski, and Wallon}}]{Anikin:2011sa}
\bibinfo{author}{\bibfnamefont{I.}~\bibnamefont{Anikin}},
  \bibinfo{author}{\bibfnamefont{A.}~\bibnamefont{Besse}},
  \bibinfo{author}{\bibfnamefont{D.~{\relax Yu}.} \bibnamefont{Ivanov}},
  \bibinfo{author}{\bibfnamefont{B.}~\bibnamefont{Pire}},
  \bibinfo{author}{\bibfnamefont{L.}~\bibnamefont{Szymanowski}},
  \bibnamefont{and} \bibinfo{author}{\bibfnamefont{S.}~\bibnamefont{Wallon}},
  \bibinfo{journal}{Phys. Rev. D} \textbf{\bibinfo{volume}{84}},
  \bibinfo{pages}{054004} (\bibinfo{year}{2011}), \eprint{1105.1761}.

\bibitem[{\citenamefont{Besse et~al.}(2013)\citenamefont{Besse, Szymanowski,
  and Wallon}}]{Besse:2013muy}
\bibinfo{author}{\bibfnamefont{A.}~\bibnamefont{Besse}},
  \bibinfo{author}{\bibfnamefont{L.}~\bibnamefont{Szymanowski}},
  \bibnamefont{and} \bibinfo{author}{\bibfnamefont{S.}~\bibnamefont{Wallon}},
  \bibinfo{journal}{JHEP} \textbf{\bibinfo{volume}{11}}, \bibinfo{pages}{062}
  (\bibinfo{year}{2013}), \eprint{1302.1766}.

\bibitem[{\citenamefont{Bolognino
  et~al.}(2018{\natexlab{b}})\citenamefont{Bolognino, Celiberto, Ivanov, and
  Papa}}]{Bolognino:2018rhb}
\bibinfo{author}{\bibfnamefont{A.~D.} \bibnamefont{Bolognino}},
  \bibinfo{author}{\bibfnamefont{F.~G.} \bibnamefont{Celiberto}},
  \bibinfo{author}{\bibfnamefont{D.~{\relax Yu}.} \bibnamefont{Ivanov}},
  \bibnamefont{and} \bibinfo{author}{\bibfnamefont{A.}~\bibnamefont{Papa}},
  \bibinfo{journal}{Eur. Phys. J.} \textbf{\bibinfo{volume}{C78}},
  \bibinfo{pages}{1023} (\bibinfo{year}{2018}{\natexlab{b}}),
  \eprint{1808.02395}.

\bibitem[{\citenamefont{Bolognino
  et~al.}(2018{\natexlab{c}})\citenamefont{Bolognino, Celiberto, Ivanov, and
  Papa}}]{Bolognino:2018mlw}
\bibinfo{author}{\bibfnamefont{A.~D.} \bibnamefont{Bolognino}},
  \bibinfo{author}{\bibfnamefont{F.~G.} \bibnamefont{Celiberto}},
  \bibinfo{author}{\bibfnamefont{D.~{\relax Yu}.} \bibnamefont{Ivanov}},
  \bibnamefont{and} \bibinfo{author}{\bibfnamefont{A.}~\bibnamefont{Papa}},
  \bibinfo{journal}{Frascati Phys. Ser.} \textbf{\bibinfo{volume}{67}},
  \bibinfo{pages}{76} (\bibinfo{year}{2018}{\natexlab{c}}),
  \eprint{1808.02958}.

\bibitem[{\citenamefont{Bolognino
  et~al.}(2019{\natexlab{d}})\citenamefont{Bolognino, Celiberto, Ivanov, and
  Papa}}]{Bolognino:2019bko}
\bibinfo{author}{\bibfnamefont{A.~D.} \bibnamefont{Bolognino}},
  \bibinfo{author}{\bibfnamefont{F.~G.} \bibnamefont{Celiberto}},
  \bibinfo{author}{\bibfnamefont{D.~{\relax Yu}.} \bibnamefont{Ivanov}},
  \bibnamefont{and} \bibinfo{author}{\bibfnamefont{A.}~\bibnamefont{Papa}},
  \bibinfo{journal}{Acta Phys. Polon. Supp.} \textbf{\bibinfo{volume}{12}},
  \bibinfo{pages}{891} (\bibinfo{year}{2019}{\natexlab{d}}),
  \eprint{1902.04520}.

\bibitem[{\citenamefont{Bolognino et~al.}(2020)\citenamefont{Bolognino,
  Szczurek, and Schaefer}}]{Bolognino:2019pba}
\bibinfo{author}{\bibfnamefont{A.~D.} \bibnamefont{Bolognino}},
  \bibinfo{author}{\bibfnamefont{A.}~\bibnamefont{Szczurek}}, \bibnamefont{and}
  \bibinfo{author}{\bibfnamefont{W.}~\bibnamefont{Schaefer}},
  \bibinfo{journal}{Phys. Rev. D} \textbf{\bibinfo{volume}{101}},
  \bibinfo{pages}{054041} (\bibinfo{year}{2020}), \eprint{1912.06507}.

\bibitem[{\citenamefont{Celiberto}(2019)}]{Celiberto:2019slj}
\bibinfo{author}{\bibfnamefont{F.~G.} \bibnamefont{Celiberto}},
  \bibinfo{journal}{Nuovo Cim.} \textbf{\bibinfo{volume}{C42}},
  \bibinfo{pages}{220} (\bibinfo{year}{2019}), \eprint{1912.11313}.

\bibitem[{\citenamefont{Bautista et~al.}(2016)\citenamefont{Bautista,
  Fernandez~Tellez, and Hentschinski}}]{Bautista:2016xnp}
\bibinfo{author}{\bibfnamefont{I.}~\bibnamefont{Bautista}},
  \bibinfo{author}{\bibfnamefont{A.}~\bibnamefont{Fernandez~Tellez}},
  \bibnamefont{and}
  \bibinfo{author}{\bibfnamefont{M.}~\bibnamefont{Hentschinski}},
  \bibinfo{journal}{Phys. Rev. D} \textbf{\bibinfo{volume}{94}},
  \bibinfo{pages}{054002} (\bibinfo{year}{2016}), \eprint{1607.05203}.

\bibitem[{\citenamefont{Arroyo~Garcia et~al.}(2019)\citenamefont{Arroyo~Garcia,
  Hentschinski, and Kutak}}]{Garcia:2019tne}
\bibinfo{author}{\bibfnamefont{A.}~\bibnamefont{Arroyo~Garcia}},
  \bibinfo{author}{\bibfnamefont{M.}~\bibnamefont{Hentschinski}},
  \bibnamefont{and} \bibinfo{author}{\bibfnamefont{K.}~\bibnamefont{Kutak}},
  \bibinfo{journal}{Phys. Lett. B} \textbf{\bibinfo{volume}{795}},
  \bibinfo{pages}{569} (\bibinfo{year}{2019}), \eprint{1904.04394}.

\bibitem[{\citenamefont{Hentschinski and
  Padr\'on~Molina}(2021)}]{Hentschinski:2020yfm}
\bibinfo{author}{\bibfnamefont{M.}~\bibnamefont{Hentschinski}}
  \bibnamefont{and}
  \bibinfo{author}{\bibfnamefont{E.}~\bibnamefont{Padr\'on~Molina}},
  \bibinfo{journal}{Phys. Rev. D} \textbf{\bibinfo{volume}{103}},
  \bibinfo{pages}{074008} (\bibinfo{year}{2021}), \eprint{2011.02640}.

\bibitem[{\citenamefont{Bolognino
  et~al.}(2021{\natexlab{b}})\citenamefont{Bolognino, Celiberto, Ivanov, Papa,
  Sch\"afer, and Szczurek}}]{Bolognino:2021niq}
\bibinfo{author}{\bibfnamefont{A.~D.} \bibnamefont{Bolognino}},
  \bibinfo{author}{\bibfnamefont{F.~G.} \bibnamefont{Celiberto}},
  \bibinfo{author}{\bibfnamefont{D.~{\relax Yu}.} \bibnamefont{Ivanov}},
  \bibinfo{author}{\bibfnamefont{A.}~\bibnamefont{Papa}},
  \bibinfo{author}{\bibfnamefont{W.}~\bibnamefont{Sch\"afer}},
  \bibnamefont{and} \bibinfo{author}{\bibfnamefont{A.}~\bibnamefont{Szczurek}},
  \bibinfo{journal}{Eur. Phys. J. C} \textbf{\bibinfo{volume}{81}},
  \bibinfo{pages}{846} (\bibinfo{year}{2021}{\natexlab{b}}),
  \eprint{2107.13415}.

\bibitem[{\citenamefont{Bolognino
  et~al.}(2022{\natexlab{a}})\citenamefont{Bolognino, Celiberto, Ivanov, and
  Papa}}]{Bolognino:2021gjm}
\bibinfo{author}{\bibfnamefont{A.~D.} \bibnamefont{Bolognino}},
  \bibinfo{author}{\bibfnamefont{F.~G.} \bibnamefont{Celiberto}},
  \bibinfo{author}{\bibfnamefont{D.~Y.} \bibnamefont{Ivanov}},
  \bibnamefont{and} \bibinfo{author}{\bibfnamefont{A.}~\bibnamefont{Papa}},
  \bibinfo{journal}{SciPost Phys. Proc.} \textbf{\bibinfo{volume}{8}},
  \bibinfo{pages}{089} (\bibinfo{year}{2022}{\natexlab{a}}),
  \eprint{2107.12725}.

\bibitem[{\citenamefont{Bolognino
  et~al.}(2022{\natexlab{b}})\citenamefont{Bolognino, Celiberto, Fucilla,
  Ivanov, Papa, Sch\"afer, and Szczurek}}]{Bolognino:2022uty}
\bibinfo{author}{\bibfnamefont{A.~D.} \bibnamefont{Bolognino}},
  \bibinfo{author}{\bibfnamefont{F.~G.} \bibnamefont{Celiberto}},
  \bibinfo{author}{\bibfnamefont{M.}~\bibnamefont{Fucilla}},
  \bibinfo{author}{\bibfnamefont{D.~{\relax Yu}.} \bibnamefont{Ivanov}},
  \bibinfo{author}{\bibfnamefont{A.}~\bibnamefont{Papa}},
  \bibinfo{author}{\bibfnamefont{W.}~\bibnamefont{Sch\"afer}},
  \bibnamefont{and} \bibinfo{author}{\bibfnamefont{A.}~\bibnamefont{Szczurek}},
  in \emph{\bibinfo{booktitle}{{19th International Conference on Hadron
  Spectroscopy and Structure}}} (\bibinfo{year}{2022}{\natexlab{b}}),
  \eprint{2202.02513}.

\bibitem[{\citenamefont{Celiberto}(2022{\natexlab{b}})}]{Celiberto:2022fam}
\bibinfo{author}{\bibfnamefont{F.~G.} \bibnamefont{Celiberto}}
  (\bibinfo{year}{2022}{\natexlab{b}}), \eprint{2202.04207}.

\bibitem[{\citenamefont{Kniehl et~al.}(2016)\citenamefont{Kniehl, Nefedov, and
  Saleev}}]{Kniehl:2016sap}
\bibinfo{author}{\bibfnamefont{B.~A.} \bibnamefont{Kniehl}},
  \bibinfo{author}{\bibfnamefont{M.~A.} \bibnamefont{Nefedov}},
  \bibnamefont{and} \bibinfo{author}{\bibfnamefont{V.~A.}
  \bibnamefont{Saleev}}, \bibinfo{journal}{Phys. Rev. D}
  \textbf{\bibinfo{volume}{94}}, \bibinfo{pages}{054007}
  (\bibinfo{year}{2016}), \eprint{1606.01079}.

\bibitem[{\citenamefont{Cisek et~al.}(2018)\citenamefont{Cisek, Sch\"afer, and
  Szczurek}}]{Cisek:2017ikn}
\bibinfo{author}{\bibfnamefont{A.}~\bibnamefont{Cisek}},
  \bibinfo{author}{\bibfnamefont{W.}~\bibnamefont{Sch\"afer}},
  \bibnamefont{and} \bibinfo{author}{\bibfnamefont{A.}~\bibnamefont{Szczurek}},
  \bibinfo{journal}{Phys. Rev. D} \textbf{\bibinfo{volume}{97}},
  \bibinfo{pages}{114018} (\bibinfo{year}{2018}), \eprint{1711.07366}.

\bibitem[{\citenamefont{Maciu\l{}a et~al.}(2019)\citenamefont{Maciu\l{}a,
  Szczurek, and Cisek}}]{Maciula:2018bex}
\bibinfo{author}{\bibfnamefont{R.}~\bibnamefont{Maciu\l{}a}},
  \bibinfo{author}{\bibfnamefont{A.}~\bibnamefont{Szczurek}}, \bibnamefont{and}
  \bibinfo{author}{\bibfnamefont{A.}~\bibnamefont{Cisek}},
  \bibinfo{journal}{Phys. Rev. D} \textbf{\bibinfo{volume}{99}},
  \bibinfo{pages}{054014} (\bibinfo{year}{2019}), \eprint{1810.08063}.

\bibitem[{\citenamefont{Babiarz
  et~al.}(2020{\natexlab{a}})\citenamefont{Babiarz, Pasechnik, Sch\"afer, and
  Szczurek}}]{Babiarz:2019mag}
\bibinfo{author}{\bibfnamefont{I.}~\bibnamefont{Babiarz}},
  \bibinfo{author}{\bibfnamefont{R.}~\bibnamefont{Pasechnik}},
  \bibinfo{author}{\bibfnamefont{W.}~\bibnamefont{Sch\"afer}},
  \bibnamefont{and} \bibinfo{author}{\bibfnamefont{A.}~\bibnamefont{Szczurek}},
  \bibinfo{journal}{JHEP} \textbf{\bibinfo{volume}{02}}, \bibinfo{pages}{037}
  (\bibinfo{year}{2020}{\natexlab{a}}), \eprint{1911.03403}.

\bibitem[{\citenamefont{Babiarz
  et~al.}(2020{\natexlab{b}})\citenamefont{Babiarz, Pasechnik, Sch\"afer, and
  Szczurek}}]{Babiarz:2020jkh}
\bibinfo{author}{\bibfnamefont{I.}~\bibnamefont{Babiarz}},
  \bibinfo{author}{\bibfnamefont{R.}~\bibnamefont{Pasechnik}},
  \bibinfo{author}{\bibfnamefont{W.}~\bibnamefont{Sch\"afer}},
  \bibnamefont{and} \bibinfo{author}{\bibfnamefont{A.}~\bibnamefont{Szczurek}},
  \bibinfo{journal}{JHEP} \textbf{\bibinfo{volume}{06}}, \bibinfo{pages}{101}
  (\bibinfo{year}{2020}{\natexlab{b}}), \eprint{2002.09352}.

\bibitem[{\citenamefont{Babiarz
  et~al.}(2020{\natexlab{c}})\citenamefont{Babiarz, Pasechnik, Sch\"afer, and
  Szczurek}}]{Babiarz:2020jhy}
\bibinfo{author}{\bibfnamefont{I.}~\bibnamefont{Babiarz}},
  \bibinfo{author}{\bibfnamefont{R.}~\bibnamefont{Pasechnik}},
  \bibinfo{author}{\bibfnamefont{W.}~\bibnamefont{Sch\"afer}},
  \bibnamefont{and} \bibinfo{author}{\bibfnamefont{A.}~\bibnamefont{Szczurek}},
  \bibinfo{journal}{Phys. Rev. D} \textbf{\bibinfo{volume}{102}},
  \bibinfo{pages}{114028} (\bibinfo{year}{2020}{\natexlab{c}}),
  \eprint{2008.05462}.

\bibitem[{\citenamefont{Babiarz
  et~al.}(2021{\natexlab{a}})\citenamefont{Babiarz, Sch\"afer, and
  Szczurek}}]{Babiarz:2020vep}
\bibinfo{author}{\bibfnamefont{I.}~\bibnamefont{Babiarz}},
  \bibinfo{author}{\bibfnamefont{W.}~\bibnamefont{Sch\"afer}},
  \bibnamefont{and} \bibinfo{author}{\bibfnamefont{A.}~\bibnamefont{Szczurek}},
  \bibinfo{journal}{PoS} \textbf{\bibinfo{volume}{ICHEP2020}},
  \bibinfo{pages}{449} (\bibinfo{year}{2021}{\natexlab{a}}),
  \eprint{2012.09721}.

\bibitem[{\citenamefont{Sch\"afer et~al.}(2021)\citenamefont{Sch\"afer,
  Babiarz, Pasechnik, and Szczurek}}]{Schafer:2021cat}
\bibinfo{author}{\bibfnamefont{W.}~\bibnamefont{Sch\"afer}},
  \bibinfo{author}{\bibfnamefont{I.}~\bibnamefont{Babiarz}},
  \bibinfo{author}{\bibfnamefont{R.}~\bibnamefont{Pasechnik}},
  \bibnamefont{and} \bibinfo{author}{\bibfnamefont{A.}~\bibnamefont{Szczurek}}
  (\bibinfo{year}{2021}), \eprint{2107.11661}.

\bibitem[{\citenamefont{Babiarz
  et~al.}(2021{\natexlab{b}})\citenamefont{Babiarz, Pasechnik, Sch\"afer, and
  Szczurek}}]{Babiarz:2021uvm}
\bibinfo{author}{\bibfnamefont{I.}~\bibnamefont{Babiarz}},
  \bibinfo{author}{\bibfnamefont{R.}~\bibnamefont{Pasechnik}},
  \bibinfo{author}{\bibfnamefont{W.}~\bibnamefont{Sch\"afer}},
  \bibnamefont{and} \bibinfo{author}{\bibfnamefont{A.}~\bibnamefont{Szczurek}}
  (\bibinfo{year}{2021}{\natexlab{b}}), \eprint{2107.14482}.

\bibitem[{\citenamefont{Cisek et~al.}(2022)\citenamefont{Cisek, Sch\"afer, and
  Szczurek}}]{Cisek:2022uqx}
\bibinfo{author}{\bibfnamefont{A.}~\bibnamefont{Cisek}},
  \bibinfo{author}{\bibfnamefont{W.}~\bibnamefont{Sch\"afer}},
  \bibnamefont{and} \bibinfo{author}{\bibfnamefont{A.}~\bibnamefont{Szczurek}}
  (\bibinfo{year}{2022}), \eprint{2203.07827}.

\bibitem[{\citenamefont{Abdul~Khalek et~al.}(2021)}]{AbdulKhalek:2021gbh}
\bibinfo{author}{\bibfnamefont{R.}~\bibnamefont{Abdul~Khalek}}
  \bibnamefont{et~al.} (\bibinfo{year}{2021}), \eprint{2103.05419}.

\bibitem[{\citenamefont{Abdul~Khalek et~al.}(2022)}]{Khalek:2022bzd}
\bibinfo{author}{\bibfnamefont{R.}~\bibnamefont{Abdul~Khalek}}
  \bibnamefont{et~al.}, in \emph{\bibinfo{booktitle}{{2022 Snowmass Summer
  Study}}} (\bibinfo{year}{2022}), \eprint{2203.13199}.

\bibitem[{\citenamefont{Hentschinski et~al.}(2022)}]{Hentschinski:2022xnd}
\bibinfo{author}{\bibfnamefont{M.}~\bibnamefont{Hentschinski}}
  \bibnamefont{et~al.}, in \emph{\bibinfo{booktitle}{{2022 Snowmass Summer
  Study}}} (\bibinfo{year}{2022}), \eprint{2203.08129}.

\bibitem[{\citenamefont{Nemchik et~al.}(1998)\citenamefont{Nemchik, Nikolaev,
  Predazzi, Zakharov, and Zoller}}]{Nemchik:1997xb}
\bibinfo{author}{\bibfnamefont{J.}~\bibnamefont{Nemchik}},
  \bibinfo{author}{\bibfnamefont{N.~N.} \bibnamefont{Nikolaev}},
  \bibinfo{author}{\bibfnamefont{E.}~\bibnamefont{Predazzi}},
  \bibinfo{author}{\bibfnamefont{B.~G.} \bibnamefont{Zakharov}},
  \bibnamefont{and} \bibinfo{author}{\bibfnamefont{V.~R.}
  \bibnamefont{Zoller}}, \bibinfo{journal}{J. Exp. Theor. Phys.}
  \textbf{\bibinfo{volume}{86}}, \bibinfo{pages}{1054} (\bibinfo{year}{1998}),
  \eprint{hep-ph/9712469}.

\bibitem[{\citenamefont{Acosta et~al.}(2022)\citenamefont{Acosta, Barberis,
  Hurley, Li, Colin, Wood, and Zuo}}]{Acosta:2022ejc}
\bibinfo{author}{\bibfnamefont{D.}~\bibnamefont{Acosta}},
  \bibinfo{author}{\bibfnamefont{E.}~\bibnamefont{Barberis}},
  \bibinfo{author}{\bibfnamefont{N.}~\bibnamefont{Hurley}},
  \bibinfo{author}{\bibfnamefont{W.}~\bibnamefont{Li}},
  \bibinfo{author}{\bibfnamefont{O.~M.} \bibnamefont{Colin}},
  \bibinfo{author}{\bibfnamefont{D.}~\bibnamefont{Wood}}, \bibnamefont{and}
  \bibinfo{author}{\bibfnamefont{X.}~\bibnamefont{Zuo}}, in
  \emph{\bibinfo{booktitle}{{2022 Snowmass Summer Study}}}
  (\bibinfo{year}{2022}), \eprint{2203.06258}.

\bibitem[{\citenamefont{Arbuzov et~al.}(2021)}]{Arbuzov:2020cqg}
\bibinfo{author}{\bibfnamefont{A.}~\bibnamefont{Arbuzov}} \bibnamefont{et~al.},
  \bibinfo{journal}{Prog. Part. Nucl. Phys.} \textbf{\bibinfo{volume}{119}},
  \bibinfo{pages}{103858} (\bibinfo{year}{2021}), \eprint{2011.15005}.

\bibitem[{\citenamefont{Anchordoqui et~al.}(2022)}]{Anchordoqui:2021ghd}
\bibinfo{author}{\bibfnamefont{L.~A.} \bibnamefont{Anchordoqui}}
  \bibnamefont{et~al.}, \bibinfo{journal}{Phys. Rept.}
  \textbf{\bibinfo{volume}{968}}, \bibinfo{pages}{1} (\bibinfo{year}{2022}),
  \eprint{2109.10905}.

\bibitem[{\citenamefont{Feng et~al.}(2022)}]{Feng:2022inv}
\bibinfo{author}{\bibfnamefont{J.~L.} \bibnamefont{Feng}} \bibnamefont{et~al.}
  (\bibinfo{year}{2022}), \eprint{2203.05090}.

\bibitem[{\citenamefont{Celiberto}(2022{\natexlab{c}})}]{Celiberto:2022rfj}
\bibinfo{author}{\bibfnamefont{F.~G.} \bibnamefont{Celiberto}},
  \bibinfo{journal}{Phys. Rev. D} \textbf{\bibinfo{volume}{105}},
  \bibinfo{pages}{114008} (\bibinfo{year}{2022}{\natexlab{c}}),
  \eprint{2204.06497}.

\bibitem[{\citenamefont{Chapon et~al.}(2022)}]{Chapon:2020heu}
\bibinfo{author}{\bibfnamefont{E.}~\bibnamefont{Chapon}} \bibnamefont{et~al.},
  \bibinfo{journal}{Prog. Part. Nucl. Phys.} \textbf{\bibinfo{volume}{122}},
  \bibinfo{pages}{103906} (\bibinfo{year}{2022}), \eprint{2012.14161}.

\bibitem[{\citenamefont{Amoroso et~al.}(2022)}]{Amoroso:2022eow}
\bibinfo{author}{\bibfnamefont{S.}~\bibnamefont{Amoroso}} \bibnamefont{et~al.},
  in \emph{\bibinfo{booktitle}{{2022 Snowmass Summer Study}}}
  (\bibinfo{year}{2022}), \eprint{2203.13923}.

\end{thebibliography}

\end{document}